%% file: arxiv-main.tex
\documentclass[conference]{IEEEtran}
\usepackage{multirow}
\usepackage{graphicx}
\usepackage{cite}
\usepackage{pifont}
\usepackage{graphicx}
\usepackage{caption}
\graphicspath{ {images/} }
\usepackage{epsfig}
\usepackage{hyperref}
\usepackage{subcaption}
\usepackage{algorithmicx}
\usepackage{algorithm}
\usepackage{algpseudocode}
\usepackage{comment}
\usepackage{listings}
\usepackage{amsfonts}
\usepackage{color}

\def\authnotes{1}
\newcommand{\figfactor}{0.48}

\newcommand{\notecolor}{blue}
\newcommand{\thenote}{\thesection.\arabic{mynote}}
\newcounter{mynote}[section]
\newcommand{\qnote}[1]{\ifnum\authnotes=1\refstepcounter{mynote}{\bf          
\textcolor{\notecolor}{$\ll$QC~\thenote: {\sf #1}$\gg$}}\fi}

\newcommand*{\ARXIV}{}

\begin{document}

%Here goes the title
\title{Logical Peering for \\ Interdomain Networking on Testbeds}

%Authors List

%\author
%{\IEEEauthorblockN{Paul Ruth, Mert Cevik, Cong Wang}
%\IEEEauthorblockA{RENCI\\
%\{pruth, mert, cwang\}@renci.org }
%\and
%https://www.overleaf.com/project/5c3cf06fe5e907418b690de4\IEEEauthorblockN{Yuanjun Yao, Qiang Cao, Rubens Farias, Jeff Chase, Victor Orlikowski}
%\IEEEauthorblockA{Duke University \\
%\{yjyao, qiangcao, rubens, chase, vjo\}@cs.duke.edu }
%\and
%\IEEEauthorblockN{Nick Buraglio}
%\IEEEauthorblockA{ESnet \\
%\ buraglio@es.net }
%}

\author
{\IEEEauthorblockN{Yuanjun Yao, Qiang Cao, Paul Ruth, Mert Cevik, Cong Wang, and Jeff Chase}
\IEEEauthorblockA{Duke University and RENCI }
%\{yjyao, qiangcao,  chase\}@cs.duke.edu, \{pruth@renci.org }
}
%\and
%\IEEEauthorblockN{Paul Ruth, Mert Cevik, Cong Wang}
%\IEEEauthorblockA{RENCI\\
%\{pruth, mert, cwang\}@renci.org }
%\and
%\IEEEauthorblockN{Nick Buraglio}
%\IEEEauthorblockA{ESnet}
%buraglio@es.net }

\maketitle

%Main body starts

\begin{abstract}
Research testbed fabrics have potential to support long-lived, evolving, interdomain experiments, including opt-in application traffic across multiple campuses and edge sites.
We propose abstractions and security infrastructure to facilitate multi-domain networking, and a reusable controller toolkit (ExoPlex) for network service providers (NSPs) running in testbed-hosted virtual network slices.  We demonstrate the idea on the ExoGENI testbed, which allows slices to interconnect and exchange traffic over peering links by mutual consent.   

%The ExoPlex prototype runs on the ExoGENI testbed, but its approach generalizes to other testbeds with similar capabilities. .  

%Our work addresses the larger challenge of linking testbeds and slices together to form larger assemblages with broader reach, including security-managed linkages into campus networks, e.g., for customer subnet cloudbursting or managed opt-in to experimental network transports.

Each ExoPlex NSP runs a peering controller that manages its interactions with its linked peers and controls the NSP's dataplane network via SDN.
Our approach expresses policies for secure peering and routing in a declarative language---logical peering.  The prototype uses logic rules to verify IP prefix ownership, filter and validate route advertisements, and implement user-specified policies for connectivity and path control in networks with multiple transit NSPs.  
\end{abstract}

\begin {IEEEkeywords}
Networks, Testbeds, SDN, Policy Based Routing, Secure Routing, Internet Security
\end{IEEEkeywords}

\section{Introduction}

\input{sections/intro}

\section{Logical Peering in ExoPlex}

\input{sections/background}

\section{Design Overview}
\input{sections/exoplex}

\ifdefined\ARXIV\
\input{sections/routeprocess}

\section{Implementation}
\label{sec:impl}

\input{sections/impl}

\fi

\section{Experiments}
\input{sections/experiment}

%\section{Related Work}
%\input{sections/related}

\section{Conclusion}
\input{sections/conclusion}

\bibliographystyle{IEEEtran}
\bibliography{main,cloud}

\end{document}

%% file: sections/intro.tex
Advanced network testbeds can serve as platforms to pilot new network transit services in testbed slices, and evolve them under real usage experience.  This paper proposes an approach to secure inter-domain networking among testbed-hosted slices, and reports on experiments using ExoGENI~\cite{baldine12:exogeni} and the Internet-2/AL2S L2 circuit service.   Our software and results generalize to testbeds with these enabling capabilities:

\begin{itemize}

\item {\bf Dynamic slices with virtual dataplanes.}  ExoGENI defines IaaS APIs to provision network topologies and program them with software-defined networking (SDN).  In our model, slices may act as {\it Network Service Providers (NSPs)} that offer transit service for IP traffic.\footnote{We program NSP dataplanes with OpenFlow SDN, which is limited to IP.}

\item {\bf NSP peering.}  ExoGENI slices may declare {\it stitchports} and interconnect (stitch) them by mutual consent~\cite{DCC17}, e.g., at an exchange site or by AL2S circuits.  Testbed support for cross-slice stitching enables NSP slices to peer at L2 programmatically, even if they have different owners.  

\item {\bf Customer opt-in.}  A slice may peer with an NSP provider and exchange IP traffic over the link.   In addition, campus networks increasingly support SDN bypass services for authorized subnets to route/accept selected traffic through a dynamic L2 network circuit, which may link to a testbed-hosted NSP.  In this way a subnet owner may ``opt in'' to use an NSP as an alternate Internet Service Provider for selected prefixes.

\end{itemize}

%For example, users of the ExoGENI testbed can write a controller program to instantiate servers and programmable routers at multiple sites and connect them with an L2 dataplane comprising local VLAN segments and/or circuits provisioned from the Internet-2 AL2S or ESnet research fabrics.  Testbeds also increasingly enable slice interconnection.  For example,  that interconnects with other networks at L2 stitching points and multiplexes their traffic over its network dataplane.

%The rich NIaaS model of these network testbeds and research fabrics enables researchers to construct and manage long-lived network transit services with experimental value-added features.  

%NSPs are also a tool to address certain challenges for testbed federation.  As a motivating usage example, consider IP interconnection via NSPs hosted on multiple research fabrics.  {\bf Unclear.}  For example, ESnet plans a shared high-bandwidth fixed IP overlay core provisioned from the ESnet fabric and available for authorized projects, which may be testbed slices or edge subnets that leverage the overlay for high-speed transit to other destinations reachable through it.   Similarly, international experiments may require their traffic to transit multiple national research fabrics for reach and connectivity.  We propose that NSPs hosted on these fabrics can interconnect their dataplanes at L2 exchanges, enabling them to peer and route IP traffic even when the fabrics do not federate directly.

These capabilities enable experimental NSP services that can carry real user traffic across research fabrics.  For example, we envision that NSPs can offer {\it security-managed} connectivity with policy controls to enable or disable flows; impose security scanning or other NFV service chains on specified flows; protect against spoofing, hijacking, and DDoS attacks; or configure other defenses that are lacking in the public Internet.  This paper extends our previous work toward that goal~\cite{CNERT18,DCC17} with support for secure policy-based inter-domain routing among {\it transit NSPs}.  It leads us to a vision of inter-domain traffic control within a network of NSPs, which may be experimental (e.g., user-managed), elastic, dynamic, and/or restricted to certain classes of traffic, e.g., high-priority data for a specific project.  It could enable advanced network services as NSPs that weave into the fabric of the Internet over time through cycles of innovation and adoption.

This paper proposes and demonstrates policy-based interdomain NSP networking in the ExoPlex toolkit---software elements that run within testbed slices and their controllers---to build NSPs and interconnect them securely.  ExoPlex addresses common security needs for experimental interdomain networking (\S\ref{sec:logical}), including route security (\S\ref{sec:design}) with custom policies for peering, route filtering, and path control, expressed in a logical trust language.  \S\ref{sec:experiments} presents experiments.

%Our prototype (\S\ref{sec:impl}) runs on the ExoGENI federated testbed; for both platforms the prefix ``Exo'' indicates a focus on approaches to interconnect multiple testbeds securely. 

%The vSDX can interconnect customer (client) edge networks that are operated by different parties.  For example, the customers might be authorized campus subnets or other testbed slices. 

%common framework for logical trust policies and certificates to secure subnet interconnection, IP prefix ownership, and edge-to-edge routing through a single 

The contributions of this paper are to: (1) expose security control abstractions for interdomain experiments with programmable security policy (a ``testbed for trust''); (2) show how to support them in a reusable toolkit using a logical trust model; and (3) demonstrate them in proof-of-concept experiments with multiple testbed-hosted transit providers and custom policies for path control.  

{\bf Note.}  This paper is an extended version of a workshop paper at the {\it International
Workshop on Computer and Networking Experimental Research Using TestBeds (CNERT)}, May 2020.  The original content is unchanged. It is extended with marked sections containing additional implementation and algorithmic detail, and additional experimental results.  
%We contend that it is valuable to provide these abstractions as a reusable toolkit to incorporate into experimental NSPs;  ExoPlex exemplifies this approach. 

%.  We propose a security architecture for cross-domain peering among IP networks provisioned on different testbeds and fabrics, as well as on fixed edge subnets in campuses and labs. 

%% file: sections/background.tex
\begin{figure*}
\begin{center}
\includegraphics[width= 0.89\textwidth]{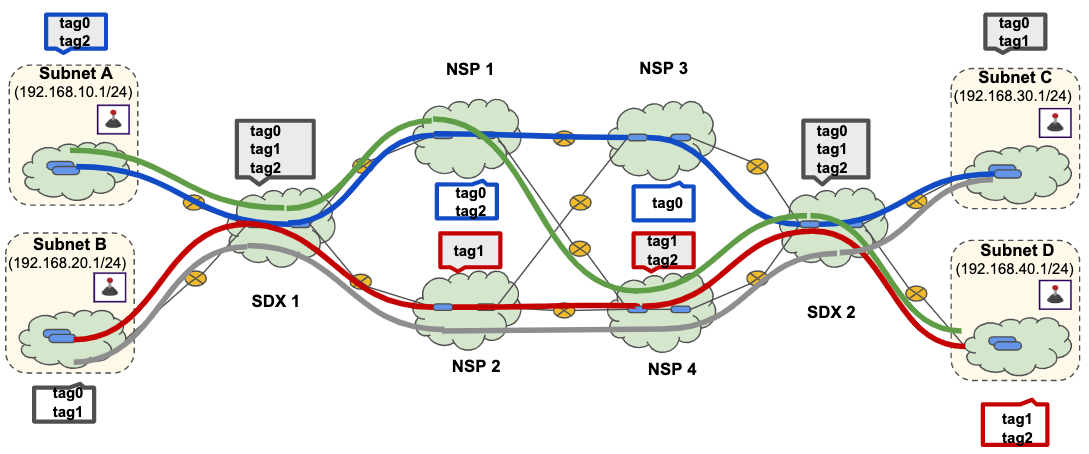}
\caption{An exemplary inter-domain network used in our experiments.  Each of the participating NSPs is instantiated as a separate ExoGENI slice with an NSP controller and a principal keypair.  The control plane comprises NSP controller APIs to establish peering links and to announce or propagate routes and security policies. Each customer edge subnet (stub) stitches to an edge provider (SDX) and specifies policies to control connectivity with other subnets and rules for packet transit, including {\it path control} policies that limit their traffic to qualifying NSPs endorsed with specified tags. For example, traffic transit between subnet A and subnet C is limited to only NSPs compatible with tag0. Similarly, B-D traffic is restricted by tag1; A-D traffic is restricted by tag2. As a result, certain flows take different paths in order to comply with each customer's traffic policies.  }
\label{fig:example-2}
\end{center}
\end{figure*}

\label{sec:logical}

%ExoPlex is a framework  to instantiate, manage, and interconnect long-lived network services (NSPs) hosted on network testbeds.

ExoPlex combines logical trust with functions for NSPs to manage an elastic network topology, control traffic in their dataplanes with SDN, and peer by stitching.  Together, these functions enable a powerful platform for policy-driven interdomain networking with a compact implementation.

{\bf The players (principals).}  Each participating network domain (NSP or edge subnet) is controlled by a security principal with a keypair.  Interacting network domains are necessarily embedded within some {\it governance} structure with additional principals, e.g., to assign addresses within a common space.  For example, communicating subnets must own compatible IP prefixes delegated to them from common trust roots, and policies may rely on security tags (attributes) of principals or networks asserted by various endorsing authorities.  Principals use their keypairs to sign their requests, delegations, policies, endorsements, and/or advertised routes.

{\bf Governance.}  ExoPlex supports an open governance model for flexible experimentation.  Each party specifies the trust roots and governance rules that it subscribes to using logic.  Parties may interact only to the extent that their structures and rules are compatible.  The experiments in this paper use a simple governance model in which common trust anchors---accepted by all participants---delegate IP prefix ownership and endorse/certify NSPs with security attributes (tags).  %ExoPlex uses the public Internet as its control network, 
Exoplex builds its secure control network over the existing public Internet,
e.g., so that NSP controllers can invoke one another's APIs for peering.

{\bf Security model for peering.}  NSP controllers expose APIs to negotiate link stitching.  An NSP's policies may limit the customers or peers that it accepts.  Once a peering link is established, either side may advertise routes for subnet prefixes to the other.  Secure interdomain routing requires that NSPs validate prefix ownership (origin authentication) and transitive route advertisements end-to-end (route validation), similarly to Internet security standards such as RPKI~\cite{rfc6810rpki} and S-BGP~\cite{lynn2000securebgp} or BGPsec~\cite{sriram2017bgpsec}.
In this paper we add customer-specified policies for off-by-default connectivity and {\it path control}, which limit traffic and constrain eligible routes based on security attributes of the NSPs and subnets.

As an exemplary demonstration, we deploy an inter-domain network with ten ExoGENI slices representing edge providers (SDX), transit NSPs, and customer domains (Figure~\ref{fig:example-2}).   In the demo scenario, customers specify path control policies that confine their traffic to compliant paths through qualified carriers---NSPs endorsed with specified tags  anchored in trust roots that the customer accepts.  For example, an endorsing authority might issue a signed assertion tagging the NSP with public key $K$ as ``production-grade safe'' or ``classified secure''.

\begin{figure*}[htb]
\begin{center}
\includegraphics[width= 0.89\textwidth]{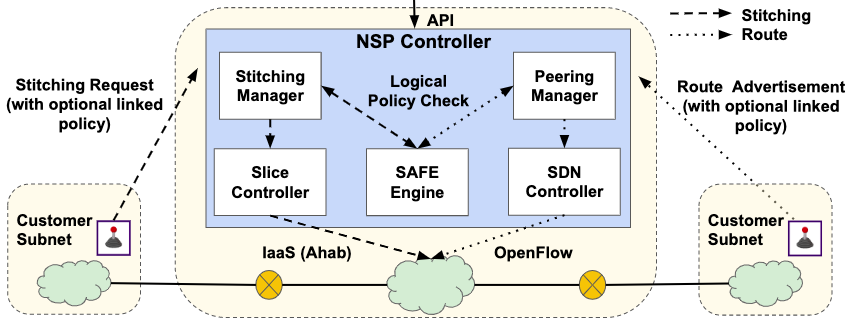}
\caption{ An exemplary ExoPlex Network Service Provider (NSP) slice.  The NSP dataplane comprises network circuits---allocated from a circuit provider such as I2-AL2S---that link one or more sites or points-of-presence (PoPs); each site runs optional NFV appliances and OpenVSwitch routers controlled via OpenFlow.   The NSP controller is a server program that invokes: testbed APIs to orchestrate the slice; OpenFlow controller(s) to manage traffic flow within the slice; and a local SAFE engine to produce, consume, and validate logical certificates and check compliance with logical policy rules.  The NSP controller exposes a northbound API for permissioned peering, permissioned flows, and policy-based path control. 
%ExoPlex is a toolkit to implement NSPs, embodying standards for logical certificates and the inter-domain control plane events.  
}
\label{fig:exoplex}
\end{center}
\end{figure*}

{\bf Standards and interoperability.}  Networks base routing and security functions on well-specified protocol standards that allow for multiple interoperable implementations.  In this work we take a first step by defining a common software platform---ExoPlex---that NSP controller software may use to manage their interactions and program their internal dataplane networks accordingly.  Logical peering in the control plane offers alternatives to relevant Internet standards (e.g, BGPsec and RPKI), but with a simpler deployment for SDN-enabled testbeds, no dataplane entanglements, flexible governance, and extended policy options (e.g., path control as in our experiments).  Because security metadata propagates through the control plane APIs over the public Internet, all crypto operations occur off of the dataplane.  

%The platform defines a minimal set of control plane APIs and a common format for extensible security certificates.  ExoPlex uses these APIs and certificates to transport all security metadata relating to IP address space management (prefix ownership), secure routing, governance and endorsement, and for security labels and policies of the NSPs and subnets that govern peering, connectivity, and path control.  

{\bf Logical trust.}  We use a logical trust language (datalog) to represent all security metadata, including endorsements, prefix delegations, routes, and policies.  The SAFE~\cite{safe:linking16} logical trust framework defines a certificate format for signed logic payloads, and a validation engine for policy checks incorporating an off-the-shelf datalog engine (Styla).   The logic vocabulary is extensible and enables a wide range of policies and trust structures without changing the certificate format or platform implementation.  Our approach is inspired by earlier work on networking using datalog, e.g., \cite{loo2009declarativep2, Sendlog:Abadi:2007}. 

The logical trust approach is a rapid prototyping vehicle for experimental approaches to secure peering.  Although the power and flexibility of trust logic imposes substantial costs as prototyped, they are off of the dataplane and accrue only on changes to the network (e.g., peer link stitching, new prefix announcements) or its security policies.  Importantly, the logic approach permits but does not require participants to write logic code: they can delegate their policies to others, take prepackaged policy off the shelf (e.g., from  federation authorities), or use packaged logic for common structures and access control abstractions.

{\bf Threat model.}  We use logical peering to express rules that defend against IP spoofing, route hijacking, and unauthorized traffic.  We provide standard logic rules for origin authentication and transitive route validation, modeling RPKI and BGPsec.  We use path control to illustrate the potential for custom policies for logical peering.  With path control, the transit path for each flow is compliant end-to-end with rules specified by the endpoints: the endpoints trust each NSP along the path to be faithful to the policies and to forward and accept traffic only along the trusted path, providing deep defenses against spoofing.  %Like other control-plane approaches, including BGPsec, ExoPlex offers no defense against an NSP that is trusted but unfaithful, i.e., that leaks traffic or that forwards or accepts traffic along an untrusted path.   

%% file: sections/exoplex.tex
\label{sec:design}

Figure~\ref{fig:exoplex} depicts an ExoPlex NSP and its controller, which is layered above its SDN controller(s), the SAFE logical trust engine, and a testbed-specific IaaS plugin (slice controller).   For ExoGENI, the IaaS plugin uses the Ahab library to build and maintain the NSP's topology by invoking ExoGENI's API for dynamic slices.  ExoPlex includes OpenFlow SDN controller software to program the NSP dataplane, based on an extended Ryu {\tt rest-router}  module.  NSP controllers may control an elastic topology or incorporate NFV and SDN-based traffic engineering.  These elements are outside the scope of this paper's focus on logical peering.

ExoPlex extends to testbeds other than ExoGENI.  NSPs may replace the IaaS plugin for another dynamic slice API, or run without one over any static SDN-programmable dataplane topology.  We have deployed ExoPlex NSPs over Corsa switch VFCs (virtual forwarding contexts) in the Chameleon~\cite{chameleon} and ESnet testbeds.

An NSP controller exposes northbound control plane APIs
\ifdefined \ARXIV
(as shown in Table~\ref{tab:apis})
\fi
for its customers and peers to request peering links and notify the NSP of new policies and routes. Calls to these APIs drive all control plane interactions to propagate routes and policies across
the  interdomain  network.  The handler for an incoming call invokes a local SAFE engine to perform various validation checks, then optionally modifies its network state and propagates notifications to peers.
Outgoing route advertisements are signed under the NSP's keypair.  NSP controllers are assumed to be reachable to one another, e.g., on the public Internet.  

 %(\S\ref{sec:sdn-network})
 
 ExoPlex includes a standard set of controller API handlers and SAFE trust scripts, which together determine when and how to install or withdraw routes and filtering rules in the NSP dataplane via the SDN controller APIs.  We extend the SDN controller for {\it ingress filtering} and {\it source-specific routes} to support policies for path control and anti-spoofing defenses.  The trust scripts define logic templates, standard validation rules for incoming routes; hooks for custom authorization rules for peer requests and permissioned flows~\cite{CNERT18}; and custom policy rules to filter outgoing routes.  We extended these rules to validate multi-hop paths through multiple transit NSPs.

\subsection{Logical Policy}

A {\it logical policy} is expressed as a set of logical facts and rules to govern and authorize routes and traffic.  NSPs subscribe to standard rules to validate routes and authenticate IP origin prefixes.   In addition, customer subnets may specify policies that guard connectivity to their prefixes and/or constrain the paths for inbound and/or outbound flows.  Associated NSPs receive those policies and evaluate compliance.   For example, a subnet's direct provider (labeled SDX in Figure~\ref{fig:example-2}) receives connectivity policy from the subnet and blocks traffic from unauthorized senders on the last hop before delivery.  
%For customer-specified path control NSPs propagate path control policies along with routes
%, by restricting the set of NSPs that may carry their traffic.  NSPs subscribe to common validation rules for the traffic that crosses their networks, and may constrain that traffic by optional local policy. 

Policy rules may query statements and security attributes of other relevant parties.  For example, connectivity rules may query attributes of the source.  The policy also defines which authorities may assert/endorse these attributes.  The standard route validation rules authenticate the origin as the owner of the prefix according to the NSP's governance rules.
%accepted by the NSP that receives the route advertisement.

The logical trust approach makes it easy to express and share governance policy in logic, independent of other elements of the implementation.  A policy might express a federation structure or, alternatively, a set of ad hoc trust agreements among the interacting parties.  For example, the prefix ownership rules in our prototype express a structure similar to the public Internet, in which prefixes are delegated transitively through a hierarchy of owners, with range containment checked at each level.  The participants must agree on the roots of authority, as in RPKI.
%, which derives ownership from a single root authority accepted by all parties. 

% An issuer of assertions or policies encodes them as logic statements in certificates signed under its keypair.

NSP controllers check policy compliance by issuing scripted queries to a local SAFE logic engine, passing a logic {\it context}---a set of certified facts and rules in datalog.  The trust scripts construct each logic context and incorporate relevant assertions and policy rules extracted from signed SAFE certificates, and selected local logic.    SAFE certificates may be passed by reference via a {\it link}, and a certificate may embed links to other certificates.  Trust scripts retrieve and follow these links to construct the context for a compliance check.  In particular, ExoPlex NSPs propagate links to customer-specified path control policies along with routes, and index them by prefix pairs in an Area-based Quad Tree (AQT)~\cite{buddhikot1999aqt}.  

%The ExoPlex SAFE package includes standard trust scripts to authorize peering and connectivity (\S\ref{sec:permissioning}), validate routes and prefix ownership (\S\ref{sec:secure-routing-overview}), and control path selection (\S\ref{sec:path-control}). 

The logic approach allows any participant to check compliance with another's policy on its behalf.   For example, customers trust their edge providers (SDX) to enforce their connectivity policies. NSPs along a valid path cooperate to enforce customer-specified path control policies; the customer trusts these NSPs to be faithful to the policy.

\subsection{Secure Routing and Prefix Ownership}
\label{sec:secure-routing-overview}

ExoPlex includes off-the-shelf trust logic scripts for secure routing, including certified route advertisements modeled on BGPsec and prefix ownership modeled on RPKI.  

%For this paper we added support for path control policies as described below (\S\ref{sec:path-control}).

{\bf Route validation.}  An NSP controller invokes a trust script to sign its route advertisements and to validate advertised routes.  Each hop of a route is a logical assertion advertising to a peer NSP a route for a specified destination prefix, along with an ordered list of predecessors (PrincipalIDs) in the path:  {\tt advertise(?DstPrefix, ?Path, ?Peer)}.  The issuing NSP invokes a script to encode the advertisement in a logical certificate and sign it under the issuer's keypair.   The certificate links to the next hop in the chain of predecessor advertisements.  

%An NSP validates a received route advertisement by invoking a script to apply logical {\tt authorizedRoute} rules (Figure~\ref{fig:safe-routing}) that precisely capture the validation criteria.  

{\bf Prefix ownership.}   The origin of a valid route must own the destination prefix.  The origin links its initial advertisement to a certificate set with evidence that it owns the prefix.  As the route propagates, each NSP in turn applies local policy rules to this logic set to validate the origin's ownership of the prefix.   ExoPlex includes a trust script to delegate a prefix to a named principal, linked to a predecessor as evidence that the issuer owns the containing prefix. 

% and logic rules to validate a chain of such delegations.  The scripts link each range

%When a customer advertises a prefix to an NSP that acts as its edge provider (SDX), the customer invokes a logic script to issue a route advertisement certificate as described above.  

%{\bf Following the model of RPKI.  How does RPKI chain range delegations?   What is the governance policy in RPKI and how to specify it?  How flexible is it?}

%{\bf Security properties.}  Secure routing in ExoPlex provides assurances similar to Secure BGP proposals.   Suppose that all NSPs follow the conventions---encoded in the shared trust scripts---to maintain and validate certified routes, and that all NSPs subscribe to the same governance model for prefix ownership.   Suppose further that NSPs install only valid compliant routes in their dataplanes and apply standard filtering as described in \S\ref{sec:sdn-network} above.   Then the inter-domain network protects the integrity of routes and defeats route hijacking and spoofed traffic.

\subsection{Path Control}
\label{sec:path-control}
\label{sec:destination_based_routing}

{\bf Path control.}  For this paper, we added support for customers (subnet owners) to express logical policy rules for {\it path control}.  These rules qualify which NSPs are eligible to carry their traffic, e.g., based on secure attributes of the NSPs.  Interdomain routing in ExoPlex finds the least-cost paths that are compliant with registered policies of both the source and destination subnets, if such paths exist.   A path (route) is compliant with the policy iff it traverses only qualified NSPs. 

%The remaining subsections (\S\ref{sec:path-control}) focus on path control.
%For path control, the governing authorities act as endorsers to label the NSPs and subnets with security attributes.    \S\ref{sec:governance} discusses governance for federation in more detail.

%Subnet owners may issue path control policies, which are sets of logic statements that say how to qualify an NSP to carry traffic to or from the subnet.   An NSP $N$ is qualified under a path control policy if a validator can prove that $N$ is qualified, given the rules in the policy and certain other statements by other parties, including endorsements of $N$'s ownership and security attributes, and related governance policies (\S\ref{sec:governance}).  

The subnet owner issues a path control policy as a certificate, and notifies its provider, passing the policy link.  A policy notification associates each policy with a prefix pair {\tt (source, dest)}, which may be wildcarded.   The route for a packet is governed by the policy with the {\it most specific} enclosing prefix pair, if any, for the packet's {\tt source, dest} addresses.   If both source and destination assert a policy, then a compliant route complies with both.

%The policy issuer publishes the policy as a logic certificate linked by a token.
%NSPs store policy links indexed by {\tt (source, dest)} prefix pairs.
%The provider SDX for the source and destination each check paths for compliance end-to-end before enabling the traffic, i.e., by installing a route and ingress rules.   Transit NSPs also check compliance of intermediate routes; these checks improve efficiency, but are not required for safety. 

{\bf Inbound path control.}  An {\it inbound} policy qualifies NSPs to carry traffic to a destination prefix, and originates from the owner of the prefix.  The subnet owner trusts the qualified NSPs, for example, to block any traffic to the destination from spoofed source addresses.  Inbound path control policies are attached to a route advertisement and propagate with the route advertisement.   Each NSP propagates routes only to peer NSPs that are compliant with the route's policy.

%A few paragraphs on outbound policies.
\input{sections/pbr.tex}

{\bf Prototype source code.}  The source code for ExoPlex and its SDN controller (in Python) is available at \url{ https://github.com/RENCI-NRIG/CICI-SAFE}, which links to a separate repository for SAFE logical trust.  The core modules of the ExoPlex NSP controller toolkit comprise about 6K lines of Java code and a few hundred lines of SAFE trust scripts.

%% file: sections/pbr.tex
{\bf Outbound path control.}  An {\it outbound} policy qualifies NSPs to carry traffic from the source prefix $S$ (whose owner specifies the policy) to the destination prefix $D$.  A customer passes outbound path control policies to the provider in a separate API call, which it may invoke at any time. 
%[{\bf currently no way to supersede old policies.}] 

Upon receiving an outbound policy event, an NSP $N$ validates its default route for $(S, D)$ (if any) for compliance with the new policy.   If the route is not compliant, then $N$ must find an alternative compliant route, even if it is longer than the current route, and then propagate it.
%to compliant peers in the usual way.
%({\bf routes?})

To do this, $N$ considers other cached routes to $D$.  Consider a cached route $R$.  $N$ received $R$ previously from a peer, but $N$ did not select or propagate $R$ because $N$ instead selected a shorter route (e.g., the current route).  If $R$ is the shortest known compliant route, then $N$ selects $R$ for $(S,D)$, replacing the current route for any flows that are within the scope of the new policy.  If $N$ knows no compliant route $R$, then it propagates the policy to at least all compliant peers that have advertised a valid route to $D$, indicating that the peer is also compliant with $D$'s inbound policy.  These peers handle the event similarly.

Eventually, if a compliant path exists, some compliant NSP identifies a compliant $R$ and advertises it as described above.  The route propagates in the usual fashion and eventually the SDX for $S$ receives it.  Along the way, each NSP on the path chooses and installs a compliant sub-route $R$ for matching flows.  If an NSP later learns of a shorter compliant route it replaces the old route in the usual way.

{\bf Policy conflicts.} A route must comply with both the outbound policy of the source and the inbound policy of the destination.  Conflicts are not a concern, although restrictive policies might block traffic entirely. If one subnet owner publishes conflicting policies for different prefix pairs, then the longest prefix match takes priority.  By convention, the source prefix dominates for an outbound policy and the destination prefix dominates for an inbound policy.

%GOOD STUFF for the long version
\ifdefined\ARXIV
{\bf Proof sketch for liveness.}  The following conditions assure discovery of a compliant path, if one exists. First, each NSP that receives the outbound policy and has no compliant $R$ propagates the policy to all compliant peers. Second, all NSPs advertise each locally selected route to all compliant peers. Lastly,
an NSP that has received the policy and knows a compliant route selects the route, and therefore propagates it to all compliant peers.  Since a compliant path can never traverse a non-compliant peer, it is sufficient to propagate the policies and routes only among compliant NSPs.  The liveness property can be proven by contradiction.

\fi

%% file: sections/routeprocess.tex
\subsection{Policy Composition and Priority}

{\it Extended content.}
%The following subsections provide more detail on how an NSP propagates routes and policies and how it indexes them by their $(S,D)$ source-destination prefix pairs.
ExoPlex routes and policies are specified to match prefixes or source-destination $(S,D)$ prefix pairs; arbitrary ranges are not supported.
Because it enables policies based on source address, it requires source-specific routing in the dataplane, and the NSP controller tracks policies and routes by $(S,D)$ prefix pairs, with optional wildcarding in either dimension.
%We evaluate these costs in \S\ref{sec:aqt-cost} and \S\ref{sec:source-specific}.  Here we present data structures and algorithms to manage such policies in the NSP controller.
For any given set of routes and policies, the NSP must select a minimal set of source-specific routes to install in its SDN dataplane.   Each installed route matches an $(S,D)$ prefix pair, and complies with policies applicable to $S$ and $D$.   Thus a route may correspond to a region of overlap among multiple controlling policies.
This section summarizes policy scope, overlap, and priority.

%This turns out not to be true: they could be smaller, as stated below.
%The policy certificates may specify these policies for larger covering prefixes or pairs, but not for arbitrary ranges. 

\begin{figure}[htb] 
\begin{center}
  \includegraphics[width=0.48\textwidth]{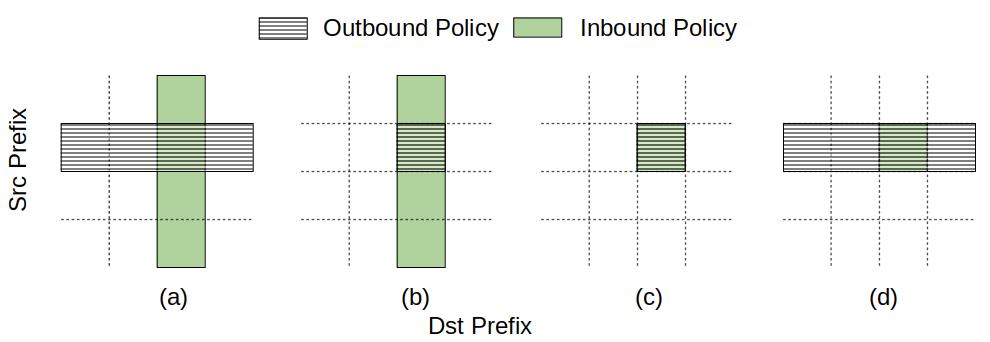} 
  \caption{Cases for two overlapping prefix pairs: they intersect, one pair contains the other, or they are equal. 
  }
\label{fig:prefixintersection}
\end{center}
 \end{figure}

Figure~\ref{fig:prefixintersection} shows how two prefix pairs can overlap.  A  prefix pair $(S,D)$ corresponds to a rectangle (a {\it region}) in the 2-dimensional source-destination IP address space.  
In each dimension of two $(S,D)$ prefix pairs, the prefix of one prefix pair is either a subset or a superset of the corresponding prefix in the other pair.
Thus the regions either cover or intersect as shown in Figure~\ref{fig:prefixintersection}.  

{\bf Containment and priority}  For simplicity, we require that prefix holders specify clear priority for policies that conflict.  For this reason, the only permitted form of overlap for policies of the same type (inbound or outbound) is containment---or equality.  For example, Figure~\ref{fig:policy-example} (discussed below) presents a scenario in which an endpoint subnet specifies an outbound policy, overriding a policy for the containing network.   
Additionally, an endpoint's inbound policies must match its advertised routes: an inbound policy's destination prefix must be the same or smaller (more specific) than the prefix for some advertised route. These restrictions do not constrain the policy, only its specification. 

 %To comply, the NSP may need to select and install a more specific routing rule to match the covered traffic.  

%For an inbound policy of the destination and an outbound policy of the source whose prefix pair overlap, the NSP need to choose a route that is compliant to both the inbound policy and the outbound policy. 

%There are cases where there are overlaps in the source prefixes of different outbound policies or overlap in the destination prefixes of different inbound policies.
These containment properties simplify policy handling.  Containment offers a clear priority rule: the more specific policy dominates. If an inbound or outbound policy is the highest priority of its type for some region, we say that the policy {\it controls} the region.   Each point has exactly one controlling policy of each type; if no policy is specified, then the default is to accept any valid route.  An NSP controller installs SDN (OpenFlow) routing rules matching each policy region, specifying the rule's priority as the area of the matching region.  In this way the most specific policy applicable to a given packet determines its route.
%Each such policy is issued by an endpoint owner who is authoritative for policy of that type in the region. 

\if 0
For multiple policies matching prefix pairs at different granularity, we apply the following priority rules:

\begin{enumerate}
	\item Inbound policies with smaller (more specific) destination prefix have higher priority.
	\item For inbound policies with same destination prefix but different source prefixes, the policy with smaller source prefix has higher priority.
	\item Outbound policies with smaller source prefix have higher priority.
	\item For outbound policies with same source prefix, the policy with smaller destination prefix has higher priority.
\end{enumerate}
\fi 

%Priority ensures that customer networks and NSPs enforce routing policies consistently.

Policy priority does not limit the flexibility of the policies.  A network owner might limit its subnets to comply with parent policies, so that more specific policies are more restrictive.  However, ExoPlex does not enforce such constraints.  We leave it to network authorities to enforce compliance by their delegates at their discretion. 
\if 0
%redundant now, but spells it out
We propose two restrictions on specifying inbound and outbound policies to allow Algorithm~\ref{alg:outbound} and Algorithm~\ref{algo:route} process route and policy advertisements correctly:

\begin{enumerate}
	\item Two inbound policies and two outbound policies SHOULD NOT intersect like in Figure~\ref{fig:prefixintersection}(a).
	\item The destination prefixes in the inbound policies SHOULD be the same as or more specific than the prefixes advertised in routes. {\bf otherwise, the inbound policy might not be effective, since in our algorithms, we require the destination prefix of the feasible route must be equal or larger than the destination prefix of the region.}
\end{enumerate}
\fi

{\bf Composing inbound and outbound policies}  Policy composition occurs when an inbound policy and outbound policy match overlapping regions.  Suppose a region $R$ is an area of overlap between an inbound policy and an 
outbound policy that both control $R$.   Then compliance may require the NSP to select and propagate a different route for traffic matching $R$ than it selects for the other regions that these policies control.   Specifically, a route for packets matching $R$ must comply with both the inbound and outbound policies that control the region.  We refer to a pair of inbound and outbound policies that control the same region as a {\it policy pair}.  Each point is governed by exactly one policy pair.

%{\bf Does a more specific subnet have to specify the campus rules?  Why not allow campus and subnet to specify them separately?}

%Since the CS subnet is managed under the authority of the campus network, the owner of CS subnet is obligated to define such policies that do not conflict with the campus network policies, given the priority rules.
%For example, when the policies of the campus network have higher priority (i.e., they must be enforced by the CS subnet), the CS subnet can import the policies of campus network in its policies. 

\subsection{Processing Advertised Routes and Policies} 

{\it Extended content.}
%An NSP receives outbound policies from policy advertisements and inbound policies attached to route advertisements.
 This section outlines data structures and algorithms to manage policies indexed by $(S,D)$ region in the NSP controller.
 Each NSP maintains a catalog containing all of the routes and policies that it knows, indexed by region in an AQT.  When it receives a new route or policy, it queries its catalog to determine adjustments to its current routes, and how to propagate the policy and affected routes to its peers.
%When the same overlapped area is covered by different inbound policies and outbound policies, the NSP chooses the inbound policy and outbound policy with the highest priority.
%For outbound policies and inbound policies whose prefix pairs overlap, the NSP need to choose a route that is compliant to both the inbound policy and outbound policy for the overlapped area. 
For simplicity, we discuss propagation of policies and routes separately. 
%Note that we may link a route to a matching inbound policy and propagate them together as an optimization. 
%We consider five events: new policy, new route, retract policy, retract route, policy update.   

\begin{table}[htb]
\begin{center}
\begin{tabular}{|p{1.8cm}|p{5.9cm}| }
 \hline
$inbound$ & AQT store for inbound policies indexed by region.\\
\hline
$outbound$ & AQT store for outbound policies indexed by region.\\
\hline
$match$ & AQT store for overlapping regions of controlling policy pairs.\\
\hline
$routes$ & Store of all accepted routes from neighbors.\\
\hline
$forwardMap$& Map of current chosen routes and their regions.  It is similar to forwarding information base (FIB), but at NSP level.\\
\hline
$exports$ & Set of routes exported to compliant neighbors.\\
\hline
\end{tabular}
\end{center}
\caption{NSP controller data structures used by Algorithm~\ref{alg:outbound} and Algorithm~\ref{algo:route}.}
\label{tab:notations}
\end{table}

{\bf Route and policy matching}. Area-based Quad Tree (AQT)~\cite{buddhikot1999aqt} supports efficient indexing of regions specified by prefix pairs. 
The root of the AQT represents the entire 2-D address space. Each AQT node has four children that equally split the parent's address space, adding one address bit in each dimension.  Thus the nodes of the AQT correspond to progressively finer-grained squares in the 2-D space.  In \cite{buddhikot1999aqt} a region or prefix pair stored in the AQT is also called a {\it filter}.  The AQT stores each filter in the highest-level node (closest to the root) that shares the same source or destination prefix, whichever dimension of the region is larger (less specific).  As the paper explains, each region is a {\it crossing filter} for the node it occupies: the region is either identical to the node's square in the 2-D space, or the region exactly crosses the square in exactly one dimension.  The filters stored at a node are the node's {\it crossing filter set}.  

Each populated node of the AQT has two collections to index its crossing filter set, one for each dimension.  It stores crossing filters in the collection that indexes the smaller (less specific) dimension of the filter.    We
choose a binary tree to represent each collection, where each node in the binary tree represents a prefix.  To insert or remove a filter, the AQT walks from the root to the target node and updates the node.  To query a prefix pair, the AQT returns all stored filters that overlap with the pair's region, leaving any conflict resolution to the routing management module.

For $N$ prefix pairs, AQT requires $O(N)$ space, $O(W)$ update (insert/delete) time, and $O(K)$ query time, where $W$ is the maximum prefix length---32 for IPv4 prefixes, and $K$ is the size of the result.

{\bf Algorithm overview: new policy}. At any given time there is a set $\mathbb{P}$ of filters of a first type (inbound or outbound), a set $\mathbb{R}$ of filters of the other type, and a set $\mathbb{M}$ of {\it match} regions.  Each set $\mathbb{P}, \mathbb{R}, \mathbb{M}$ is indexed in its own AQT.

Different policy pairs may match on the same region, but each match is controlled by exactly one most specific policy pair:
$\forall m \in \mathbb{M},\ \exists p_m\in \mathbb{P},\ \exists r_m\in \mathbb{R},\ m = p_m \cap r_m \land
%for all $m$ in $\mathbb{M}$, there exists $p_i$ in $\mathbb{P}$ and $r_i$ in $\mathbb{R}$ such that $m = p_i \cap r_i$. And $p_i$ and $r_i$ are the most specific (smallest) filter 
\forall p\in \left \{p\in \mathbb{P} \mid m \subseteq p \right \}, p_m \subseteq p
\land \forall r\in \left \{r\in \mathbb{R} \mid m \subseteq r \right \}, r_m \subseteq r$.
%{\bf Correctness.}
{\it Proof}.  Let $\mathbb{X}$ be the set of all points in 2-D address space. Because of the containment property, there is a most specific filter $p_x$ in $\mathbb{P}$ and a most specific filter $r_x$ in $\mathbb{R}$ that control each point $x$.  The most specific filter may be the default filter.  
The most specific policy pair that controls $x$ is then $(p_x, r_x)$.  The corresponding match region $m_x = p_x \cap r_x$ is the minimal match region that controls $x$: 
$\forall x \in \mathbb{X},\ \land\ \forall m \in \left\{ m \in \mathbb{M} \mid x \in m \right \},\ m_x \subseteq m$.

%The algorithms are correct iff 
%the minimum region that contains $x$ is controlled by $p_x$ and $r_x$: 

Consider addition of a new filter $p$ of the first type.   Existing matches $m \in M$ that do not overlap $p$ ($m \cap p$ is empty) are unaffected.   If $p \in P$ then for each $r \in R$ that overlaps $p$ the algorithm must visit the existing match $m = p \cap r$ to determine if it should use the new policy.   If $p \notin P$ then each such $r \in R$ may create a new match region $m = p \cap r$ to add to $M$.

Whether or not $p \in P$, there may exist one or more matches $m$ that overlap: $m = p_i \cap r_j$ with $p_i \in P, r_j \in R, p_i \neq p$, $m \in M$, and non-empty $p \cap m$.   Because of the containment property, for any such $m$, there are exactly two cases.  Case 1: $p_i \subset p$.   Then $p_i$ dominates $p$, so the new $p$ does not affect this $m$.  Case 2: $p \subset p_i$.   Then the new $p$ introduces a match filter $m' = p \cap r_j \subseteq m$ that dominates $m$ for sub-region $m'$.  Filter $m$ is unaffected and remains in place to control the rest of its region.  It is possible that $m' = m$ if $r_j$ itself is more specific than $p_i$ in at least one dimension, but $p$ updates policy for $m'$ regardless.

%{\bf (m' is not necessarily new. If $r_j$ is more specific than both $p$ and $p_i$, the match will be the same, though we need to update the policy pair for the match (the latter case in line 7 of Algo 1). When the new match is smaller and new, that is the former case of line 7 in Algo 1)}.  The new filter , but 

The algorithm handles all of these cases by considering $m = p \cap r$ for each overlapping $r \in R$ in priority order where such order exists.   If $m \notin M$, then add $m$ to $M$ controlled by policy pair $(p, r)$.   Else $m \in M$ and $m= p_i \cap r_j$ for unique $p_i$ and $r_j$ as described above.  If $p_i \subset p$ then there is no change for this $m$ because $p_i$ dominates the new $p$ in this region.  If $p_i = p$ or $p \subset p_i$, then the new policy supersedes the old one in $m$.  However, if $r_j \subset r$ then there is no need to update $m$: $r_j$ has higher priority than $r$, so the algorithm has already considered $p \cap r_j$ and so has already added or updated $m$ for $p$.

{\bf Example: new outbound policy}. Algorithm~\ref{alg:outbound} shows the NSP procedure to process a received outbound policy $policy_{ob}$ with the specified region and policy certificate. The NSP stores the outbound policy in $outbound$ AQT indexed by its region.
Then it runs a query to the $inbound$ AQT to retrieve a list of all applicable inbound policies that overlap with $policy_{ob}$ (line 3), in descending order of policy priority (ascending order of region area).
Then for each inbound policy $policy_{ib}$ in the list, it computes the overlapped region of the inbound policy and the new outbound policy.  It then queries the $match$ AQT for this region (line 6).

If the region is not present in $match$, then add it for this policy pair: the matched inbound policy $policy_{ib}$ and the new policy $policy_{ob}$ control the region.
Suppose instead that there exists a policy pair in $match$ with the same overlapped region.  If $policy_{ib}$ is the same inbound policy as in the pair that is most specific for the region,
%the inbound policy of the existing pair must be the most specific inbound policies that controls the region. When receiving a new outbound policy, the most specific inbound policy for the region doesn't change. Previous, we said if $policy_{ib}$ has equal or higher priority. That was still true, but $policy_{ib}$ won't have higher priority. 
and $policy_{ob}$ has equal or higher priority (more specific) than the outbound policy of that pair, then the new policy $policy_{ob}$ controls: update the policy pair for the region (lines 7-8).
Otherwise, at least one of $policy_{ob}$ or $policy_{ib}$ is of lower priority (less specific) than the corresponding policy of the pair.  If it is $policy_{ob}$, then we do not update the region because the previous outbound policy dominates in this region.  If it is $policy_{ib}$ but not $policy_{ob}$, then we have already processed a more specific inbound policy dominating the overlapped region.

%is updated as the received outbound policy and the matched inbound policy (line 7-8). 
%{\bf Still not clear.} {\it For example, given an existing inbound policy (*, 2.2.2.0/24) and outbound policy (1.1.1.0/24, *), the overlapping region is (1.1.1.0/24, 2.2.2.0/24). Now consider a new outbound policy(1.1.1.0/24, 2.2.2.0/24), which dominates the previous outbound policy. The overlapping region is still the same (1.1.1.0/24, 2.2.2.0/24), but the new outbound policy has higher priority than the old outbound policy. So the policy pair for the region should be the inbound policy and the new outbound policy. And we update the policy pair in line 8 accordingly.  Now we consider another example, for a new inbound policy (*, 2.2.0.0/16), the overlapping region with the new outbound policy (1.1.1.0/24, 2.2.2.0/24) is the same (1.1.1.0/24, 2.2.2.0/24). But the new inbound policy has lower priority than the existing inbound policy, so we don't update for the region (1.1.1.0/24, 2.2.2.0/24).}
%If the outbound policies have the same priority, then  Specially, same priority in outbound policies means that the new outbound policy is an update to the previous outbound policy for the same prefix pair,  and the same priority in inbound policies means that the two inbound policies are the same. 
If the new $policy_{ob}$ controls any region, then find a (new) compliant route for the region.  The NSP retrieves all accepted routes whose destination prefixes fully cover the destination prefix of the overlapped region in order of preference; currently the preference order is by path length, but the NSPs are free to prefer routes by other attributes. 
If a route is the most preferred among compliant routes to the policy pair of the region, the NSP chooses the route for traffic in the overlapped region and exports the route if it has not done so already (line 11-19).
If there is no compliant route for the pair, the NSP drops any matching traffic in the overlapped region.
%The priority order avoids unnecessary exports of routes for low-priority policies when higher-priority policies are available. 

\begin{algorithm}[htb!]
\begin{flushleft}
  \caption{Process a new outbound policy.  See Table~\ref{tab:notations} for notation. The $query$ method of the AQT
  %store in the pseudo code
  returns a list of objects whose regions overlap with the queried region.
}
  \label{alg:outbound}
  \begin{algorithmic}[1]
      \State \textbf{event} $policy_{ob}(region, cert):$
      \State $outbound.put(policy_{ob}.region, policy_{ob})$
      \State $ibPolicies = inbound.query(policy_{ob}.region)$
      %\State $exportRoute  \gets \{\}$
      \For{$policy_{ib} \gets ordered(ibPolicies)$}
        \State $region_{ol} = policy_{ob}.region \cap policy_{ib}.region$ 
        \State $\langle policy_{ob}^{prev}, policy_{ib}^{prev}\rangle = match.get(region_{ol})$
        \If {$region_{ol} \notin match$ \textbf{or} $(prio(policy_{ob}) \geq prio( policy_{ob}^{prev})$ 
        \textbf{and} $policy_{ib} == policy_{ib}^{prev})$}
          \State $match.put(region_{ol}, \langle policy_{ob}, policy_{ib} \rangle)$
          \State $orderedRoutes = routes.filter(r \rightarrow  region_{ol}.dst \subseteq r.dst)$
          \State $bestRoute = null$
          \For {$route \gets orderedRoutes$}
            \If{$compliant(route, policy_{ib}, policy_{ob})$}
              \State $bestRoute = route$
              \If{$route \notin exports$}
                 \State $exports.add(route)$
              \EndIf
              \State \textbf{break}
            \EndIf
          \EndFor
          \State $forwardMap.put(region_{ol}, bestRoute)$
        \EndIf
      \EndFor
   % \EndProcedure
  \end{algorithmic}
\end{flushleft}
\end{algorithm}

\begin{figure}[htb!] 
\begin{center}
  \includegraphics[width=0.45\textwidth]{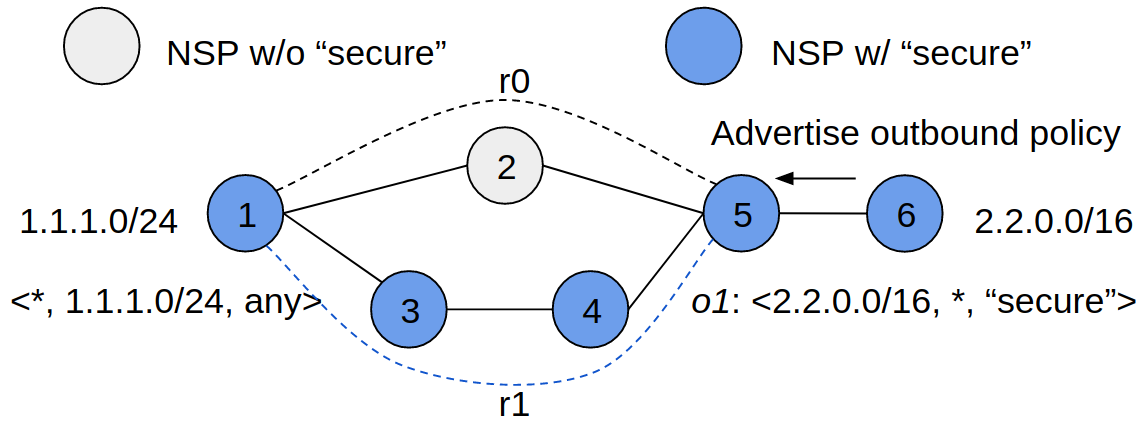} 
  \caption{An example scenario in which policy overlap forces selection of an alternate route for the intersection.  NSP 5 processes a new outbound policy $o_1$ from 6 that requires NSPs with attribute ``secure'' for its outbound traffic, forcing matching traffic to the longer path (in blue).}
\label{fig:policy-example}
\end{center}
 \end{figure}

Figure~\ref{fig:policy-example} shows an example scenario in which NSP 5 processes a new outbound policies from NSP 6, which requires a different path for the specified region. Suppose that NSP 5 have learned all routes and policies except the outbound policy from 6 at the beginning. Then 6 advertises its outbound policy $o_1$ and NSP 5 chooses a different path for the region accordingly.  Table~\ref{tab:process-policy-example} shows the states of NSP 5 at different stages.   
%advertises a new path that is compliant to both the inbound and outbound policies for the overlapped prefix pair.

\begin{table*}[htb!]
\begin{center}
\resizebox{\textwidth}{!}{%
\begin{tabular}{|l|l|l|l|l|l|l|}
 \hline
event&routes & inbound & outbound & match& forwardMap & exports\\
\hline
\multirow{2}{*}{start}
&$r_0:$
$1.1.1.0/24, [2,1]$
& \multirow{2}{*}{$i_0:
\langle *, 1.1.1.0/24, any\rangle$} 
&\multirow{2}{*}{$o_0:\langle *, *,  any \rangle$ }
&\multirow {2}{*}{$\langle *, 1.1.1.0/24 \rangle \rightarrow \langle i_0, o_0 \rangle$}
&\multirow{2}{*}{$\langle *, 1.1.1.0/24 \rangle \rightarrow r_0$}
&\multirow{2}{*}{$r_0$}\\
&$r_1:$$1.1.1.0/24, [4,3,1]$ &&&&&\\
\hline
\multirow{2}{*}{$o_1$}
&$r_0: 1.1.1.0/24, [4,3,1]$
&\multirow{2}{*}{$i_0:\langle *, 1.1.1.0/24, any \rangle$}
& $o_0:\langle *, *, any\rangle$
&$\langle *, 1.1.1.0/24 \rangle \rightarrow \langle i_0, o_0 \rangle$
& $\langle *, 1.1.1.0/24 \rangle \rightarrow r_0 $ 
&$r_0$\\
&$r_1: 1.1.1.0/24, [2,1]$ & &
$o_1: \langle 2.2.0.0/16, * ,``secure" \rangle$
& 
$\langle 2.2.0.0/16, 1.1.1.0/24 \rangle \rightarrow \langle i_0, o_1 \rangle$ 
&$\langle 2.2.0.0/16, 1.1.1.0/24 \rangle \rightarrow r_1$
& $r_1$\\
\hline
\end{tabular}}
\end{center}
\caption{The states of NSP 5 before and after processing the outbound policy from 6 in Figure~\ref{fig:policy-example}.}
\label{tab:process-policy-example}
\end{table*}
 
{\bf New inbound policy}. The procedure to process a new inbound policy is similar to Algorithm~\ref{alg:outbound}.
%except that there is a default outbound policy on all NSPs for $\langle \ast, \ast \rangle$ that allows all routes to be exported when there is no specified outbound policy for the prefix pair. If the destination doesn't specify any inbound policies for its routes, the NSP assumes that there is a default inbound policy that accepts any route to the destination subnet.

{\bf New route}. Algorithm~\ref{algo:route} shows the procedure to process a new route that the NSP accepts.  
First, the NSP retrieves all policy pairs for overlapped regions by the destination prefix. We  use the route only if the destination prefix of the overlapped region is covered by the advertised destination prefix in the route. If the new route is compliant to the policy pair and is preferred over the existing best route (which may be empty), the NSP chooses the new route for traffic matching the overlapped region and exports the route to its neighbors.

\begin{algorithm}[htb!]
\begin{flushleft}
  \caption{Process a new accepted route.}
  \label{algo:route}
  \begin{algorithmic}[1]
    \State \textbf{event} $route(dst, path):$
    \State $routes.add(route)$
    \State $policyPairs = match.query(\langle \ast, dst \rangle) $
    \For {$\langle policy_{ob}, policy_{ib}\rangle \gets policyPairs$}
      \State $region_{ol} = policy_{ob}.region \cap policy_{ib}.region$ 
      \If {$region_{ol}.dst \subseteq route.dst$}
        \State $route^{prev} = forwardMap.get(region_{ol})$
        \If {$compliant(route, policy_{ib}, policy_{ob})$ 
        \textbf{and} $pref(route) > pref(route^{prev})$}
          \State $forwardMap.put(region_{ol}, route)$
          \State $exports.add(route)$
        \EndIf
      \EndIf
    \EndFor
  \end{algorithmic}
 \end{flushleft}
\end{algorithm}

 The forwarding path of the highest-priority (most-specific) OpenFlow entry for a packet in dataplane must comply with the most specific policy pair (match region) that controls its source and destination IP address.  Algorithm~\ref{alg:outbound} ensures this property because: (1) it considers every possible match region by processing every inbound policy that overlaps with the new outbound policy; (2) it always identifies the most specific match by ordering inbound policies in descending priority; (3) it updates the policy for a match region only if the new policy is more specific than the existing policy.  Given that property, routing is correct because: (1) the controller installs an OpenFlow entry for each match region; (2) it assigns the priority of the entry as the area of the corresponding region. Thus OpenFlow entries that match on more specific (smaller) regions have higher priority in the dataplane.

\subsection{Source-specific routing with SDN-enabled dataplane}
To support inbound and outbound path control policies, we need to
implement source-specfic routing that match on both source and destination IP address in SDN-enabled dataplane. We can run OVS in ExoGENI slice to support SDN.
Testbeds like Chameleon and ESnet also deployed hardware switches  (Corsa DP2000 series) that support SDN and virtualization. 
They allow users to manage isolated networks with OpenFlow-enabled virtual forwarding contexts (VFC) and their own OpenFlow controllers. 

Source-specific routing requires more matching fields in OpenFlow entries and potentially more OpenFlow entries. We evaluate the overhead of source-specific routing in OVS in \S\ref{sec:source-specific}. The matching fields and the number of OpenFlow entries in hardware flow table do not affect the performance of hardware switches~\cite{bauer2018devicebench}, as they can match different OpenFlow entries in parallel with TCAM (ternary content-addressable memory). However, hardware switches come at higher economic cost and have limited hardware flow table capacities.

%% file: sections/impl.tex
{\it Extended content.} This section presents an extended treatment of the ExoPlex prototype. We focus on how it uses logical trust to implement a trust plane for secure interdomain networking. 

ExoPlex is designed to operate across inter-connected SDN networks (dataplanes), including virtual SDN networks hosted on testbeds.  We make minimal assumptions about the capabilities of their switch infrastructure or virtual hosting services.  Each ExoPlex NSP is under the control of a domain.  The NSPs correspond to autonomous systems (AS) in Internet BGP routing protocols.  We use a different term to avoid confusion with those protocols, since ExoPlex does not use BGP or assume any support for BGP.  Instead, it implements secure routing and security policy using authenticated logic exchanges in the control plane, as described below. 

Thus ExoPlex is compatible with a range of testbeds and SDN systems.   Its control plane and trust plane are entirely decoupled from the protocols used to operate the SDN dataplanes: it operates above the SDN networks and is unknown to them.   It is also transparent to the hosting providers (e.g., testbeds), because NSPs use the providers' native APIs to manage virtual infrastructure for the NSP dataplane.   For example, on ExoGENI, an NSP may use APIs in the Ahab toolkit to provision its nodes and links~\cite{DCC17}, and APIs for cross-slice peering to stitch NSP dataplane networks at L2 via the {\it stitchport} abstraction or via direct peering at a common hosting site~\cite{xin2016towards}.  Customer networks may be campus
subnets that connect to supported transport fabrics (e.g., I2-AL2S, ESnet) and can attach to stitchports. 
%ExoGENI slices can also connect to other networks and testbeds, e.g., through a programmable exchange at Starlight.
%affiliate with ExoGENI, i.e., they 

%Our prototype runs above the ExoGENI testbed and enables NSP slices to provide network transit service to customer slices on ExoGENI and Chameleon. 

\if 0
  ExoGENI provides full NIaaS APIs to manage and interconnect elastic NSPs.  
An NSP slice operates a slice dataplane, a virtual topology spanning multiple points of presence in ExoGENI.   
We use OpenFlow and OpenVSwitch VMs for the switches and routers, and a standard OpenFlow SDN {\it network controller} built with the Ryu controller framework.  The slice controller invokes Ahab APIs to request VMs and links from the ExoGENI testbed.   ExoGENI orchestrates end-to-end stitching of the cross-site links from 
network circuits allocated dynamically from network circuit providers such as I2-AL2S and ESnet.  
\fi

\subsection{Control Plane}

The ExoPlex control plane operates at the level of the per-domain NSP controllers.  An NSP controller is a server, operated on behalf of an NSP, that speaks for the NSP and commands its network.  The controllers interact with one another and with other servers operated on behalf of their customers.
Additionally, an NSP controller commands the NSP's dataplane---its network of SDN switches---through the northbound APIs of its SDN controller(s).  It may also call virtual hosting APIs to add or remove virtual switches or links.   We assume a separate control network for all of these control-plane interactions.  The control network is operated by infrastructure providers---campuses, cloud providers, and network testbeds---and is accessed via the public Internet.
%that host the physical infrastructure,

The NSP controllers export RPC APIs (e.g., REST/HTTP) to control inter-domain peering and networking.
%over authenticated and encrypted transport connections
Table~\ref{tab:apis} summarizes selected northbound control plane APIs for an NSP controller.  The customers and peers invoke these APIs to attach (stitch) an L2 link and to enable specified IP traffic to flow over the link.   These calls propagate the routes and policies that govern traffic flow, which are encoded in logical certificates. 

\ifdefined \ARXIV
\begin{table}[htb]
\begin{center}
\begin{tabular}{|p{2.5 cm}|p{5.3cm}|}
\hline
{\bf stitchRequest}(slice ID, sliver ID, secret, stitch properties) &  Stitch  a  sliver  (node) in  a customer or peer slice to an NSP edge node at the same site.  ExoGENI supports such cross-slice L2 stitches guarded by a {\it secret}, as in~\cite{DCC17}.\\
\hline
{\bf undoStitch}(slice ID, sliver ID) & Discard a stitched L2 link between the peer/customer sliver and the NSP slice.\\
\hline
{\bf stitchportRequest}( stitchportURL, vlan, stitch properties) &  Stitch a science network outside of GENI to the NSP slice at a static stitchport~\cite{DCC17}.\\
\hline
{\bf advertiseRoute}(route, route cert)  & Advertise a route, with a link to the signed certificate of the route.\\
\hline
{\bf advertisePolicy}(src, dst, policy cert) & Advertise  a  path  control policy  for  traffic  from  the source prefix to the destination prefix, with a link to the signed policy certificate.\\
\hline
\end{tabular}
\end{center}
\caption{Control plane APIs of an NSP controller.  Customers and peers invoke these REST APIs to attach (stitch) a node to the NSP, and to notify it of routes for a peering link and of policy rules governing the use of those routes. These requests are authenticated and authorized as described in \S\ref{sec:impl}.}
\label{tab:apis}
\end{table}
\fi

\subsection{Logical Trust}
\label{sec:logical-trust}

The NSP controllers include trust modules to check policy compliance (\S\ref{sec:nsp-authz}) as they handle API calls in Table~\ref{tab:apis}.  All trust metadata for compliance checks and other trust decisions is encoded in logical certificates passed through the calls in Table~\ref{tab:apis}.   More precisely, the parameters include {\it tokens} that reference certificates and certificate chains.   The certificates reside in a shared repository indexed by these tokens.  Certificates may also include tokens that link to other certificates.  Thus issuers may link certificates together to form DAGs and chains.

Each principal in the system possesses a keypair to sign its certificates, and a principal ID (PID) that is the hash of its public key.  In general, principals may mint their own keypairs: there is no designated PKI hierarchy.  Instead, participants use the logic to specify a governance structure to endorse keypairs (PIDs) to the extent required (e.g., see \S\ref{sec:example}).   For example, a certificate may contain a logical assertion that endorses another PID for a specified attribute or role, or delegates specified authority or ownership to it.   Each principal stores its full public key in the repository indexed by PID; thus, knowing another principal's PID is sufficient to authenticate it.    A principal may link its PID to other certificates issued by other parties, as proof of its identity, attributes, and/or privileges.

%A principal may issue a certificate signed under its keypair, select a token to name the certificate, and store it into the repository under the token.    

{\bf SAFE}.  These logical trust features are implemented in the SAFE platform.  ExoPlex is an application of SAFE.
SAFE is programmable at two levels.  Each participant in a SAFE application installs and runs trust scripts written in a simple interpreted scripting language.   Operators choose the trust scripts to install as part of a principal's trusted computing base.  The SAFE certificates are also programmable: they contain assertions (such as endorsements and delegations) and/or policy rules in a datalog logic language.  Thus SAFE certificates are essentially fragments of logic programs.   Participants exchange certificate data through the store and interpret the logic in one another's certificates.

The trust scripts run under an interpreter in a local SAFE instance.  
Operationally, the SAFE instance resides in a local process under some principal's control, running with access to its keypair.
The instance also includes an off-the-shelf datalog inference engine (Styla) for querying logic assembled from local policies and imported certificates.  Logic queries are safe and sound: they complete in bounded time, track attribution for each statement or belief, and consider only assertions by principals that are properly trusted for their content according to policy.

{\bf Trust scripts}.    Trust scripts running within a SAFE instance have no contact with the outside world except puts and gets on the certificate repository (store).  Calls to these scripts might originate from administrative commands entered by a human or from event handlers within some application or service.
   
Trust scripts include {\it issuer} scripts, which construct certificates and post them on the store, and {\it guard} scripts, which gather sets of certificates from the store and run logical queries against them to check compliance with logical policies.
An issuer may embed arbitrary logic into its certificates, generated according to parameterized logic templates in its issuer scripts.
When running a guard, a SAFE instance fetches linked certificate DAGs from the store, checks signatures and other criteria (e.g., TTLs) to validate individual certificates, extracts and caches the logical payloads, assembles cached logic into a logic program context, and invokes logical queries on the context.

% .  It is safe for them to evaluate one another's logic extracted from these certificates.  In contrast, trust scripts are trusted programs.
%

{\bf Certificate repository}.  SAFE's certificate store is suitable for decentralized operation with the trust properties of a permissioned blockchain deployment, but with a more scalable implementation.  Specifically, it is intended to run as a Byzantine quorum system (BQS) following Phalanx~\cite{phalanx}.  These systems scale more easily than blockchains because they allow sharding, in which each operation executes on only a subset of replicas.  They are sufficient for logical trust because the logic programming model  does not depend on a linear sequence of operations as imposed by blockchains, in which all operations execute on all replicas in a strict linear order for state-machine consensus.  
%We refer to SAFE's combination of trust logic and a BQS replicated store as Unchained Logic. 
Thus ``unchained logic'' offers a scalable alternative to blockchains as a foundation for decentralized trust; ExoPlex shows how to use it for secure internetworking.   However, the current ExoPlex prototype uses an enterprise key-value store (Riak) operated by a trusted party.

\subsection{Example Scenario: FabNet}
\label{sec:example}
We consider an example ExoPlex scenario: FabNet, a hypothetical secure internetwork for a community of scientists.  It comprises an assemblage of network resources spanning multiple campuses and research fabrics.    These resources are allocated and programmed for use by researchers in some field.   FabNet traffic crosses multiple network providers (NSPs) resident on those fabrics, as well as the campus networks at the edges.   A FabNet consortium approves and endorses participating campuses and NSPs.

Traffic traverses FabNet by agreement of the sender and receiver of the traffic.  The endpoints are subnets on the attached campuses.   Suppose a campus network authority (CNA) assigns an IP prefix to a secure subnet and delegates ownership and limited control of the subnet to a research group.  The research group leader issues a request to enable connectivity with a collaborator's subnet on a peer campus.   The endpoints agree that traffic between them shall traverse FabNet.

The campuses and FabNet NSPs cooperate to direct selected IP traffic through FabNet.  To attach to FabNet, a campus network operator establishes circuit connectivity to a selected FabNet NSP edge site.
They cooperate to validate all prefixes and routes to ensure that the traffic is authorized for FabNet, and that it transits only approved FabNet NSPs.   
%{\bf [How exactly does this work, pre-arranged, manual, or automated when a FabNet connection is requested..]}

{\bf FabNet governance}.  This example features multiple governance authorities and other identities interoperating with the NSPs.  
The governance structure defines a natural PKI hierarchy via logical trust.    
For example, the FabNet consortium root acts as a shared trust anchor whose PID is known to all participants.  It endorses the PIDs of the CNAs and NSPs.   The CNAs also endorse campus subnets, assert attributes of subnets, and delegate selected management authority to researchers that control those subnets.     Ownership of the containing prefixes is certified by a delegation hierarchy rooted in FabNet or some other authority, e.g., ICANN.   Section~\ref{sec:governance} discusses governance in more detail.

% The CNAs run identity services that endorse the researchers and assert attributes, such as project memberships. 

%Campus IT staff operate the CNAs, which allocate campus prefixes to subnets and delegate control of routing policy for subnets to their owners, within limits of campus policies.

\subsection{Secure Routing with Logical Trust}

The ExoPlex prototype includes SAFE trust scripts used by all participants in an ExoPlex network.  The scripts embody sample logic for the secure routing mechanisms and policies in this paper.    Specifically, they include standard datalog logic rules to secure all routing with origin authentication (following RPKI) and route authentication (following BGPsec).

The prototype also includes exemplary policy logic rules and certificate templates that endpoints may use to authorize peering, connectivity, routing (path control), and governance.  The exemplary policies are sufficiently powerful to implement FabNet and other protected networks over a set of NSPs.  Crucially, these elements do not affect the NSP controllers or their API in Table~\ref{tab:apis}.   Instead, various principals invoke issuer scripts to issue linked certificates and policy rules to validate them.  
The endpoints may define new policy types by programming new assertion types and validation rules into their trust scripts.  The NSP controllers invoke standard guard scripts that import the certificate DAGs and query them for compliance under the applicable rules.  Our approach can also accommodate NSP routing policies, e.g., to prefer certain routes or exclude unauthorized traffic, but we do not discuss NSP policies in this paper.
%{\bf Can FabNet reject traffic for unauthorized pairs?}

%{\bf [Do we have any example of a mobile policy rule in ExoPlex now?  They are expensive.]}

\if 0
%Mobile
The NSPs import these certificates and evaluate policy statements and rules issued by various endpoints; these policies together determine how the NSPs propagate and adopt the routes.   Any participant that relies on an NSP $A$ for packet transit anywhere in a path also trusts $A$ to apply that logic correctly, e.g., by using the ExoPlex code without modification. 

 It does not require changing the NSPs or their trust scripts.   Instead, the new policies and assertions are encoded in logic and transmitted as signed certificates.

%  Thus it is easy to use ExoPlex with fluid governance structures established by mutual agreement among the participants.
%In this way developers may use SAFE to define a range of trust and governance structures involving arbitrary principals, e.g., acting as authorities and trust anchors. 
\fi

The exemplary policy logic is based on Attribute-Based Access Control (ABAC).  Authorities use a common logic package for ABAC to generate attributes, assign or delegate them to other principals, and check that specified attributes are present.   The implementation represents attributes as {\it tags}.
A  tag is a string name that represents a permission, role, group, or attribute.
A logic policy may state that a principal is authorized if and only it wields or possesses the tag, or perhaps a conjunction or disjunction of tags.  For example, a subnet owner may express policies that restrict the set of NSPs that can carry its traffic, based on their tags. 
%The implementation seems to presume that each NSP knows its tags and keeps certs for them in a wallet.

Each tag has a controlling authority---a principal who creates the tag and defines the rules for delegating the tag.   Any principal may create tags and act as a tag authority.  The tag authority (or root) has sole power to specify which principals wield the tag.    In the implementation, tags are self-certifying: the tag name is a concatenation of its root's PID and a name (such as a UUID) chosen by the tag root.   An authorizer accepts a tag delegation as valid only if the tag's root asserts it or accepts it under its rules.

\subsection{Authorization for the NSP API}
\label{sec:nsp-authz}

The NSP APIs in Table~\ref{tab:apis} invoke guard scripts to check authorization before completing each call.   
Clients pass tokens for certificate chains that contain applicable policy rules and any relevant credentials, endorsements, or delegations granted to them by other parties.  The guard rules may consider any of this information in deciding whether to grant access.  Guard queries apply  
logical rules that are satisfied only if all required certifications are present and valid.   
These include certificates from various authorities endorsing principals and objects and asserting their attributes.  

{\bf Stitching}.
A stitch request from another slice passes the $sliceID$ of the requester.  The $SliceID$ can serve as a token for a certificate chain that identifies the slice, including any attributes and a binding to a project group, e.g., following the SAFE instantiation of the GENI trust structure~\cite{cao2017cloudfed, brinn2015genitrust}.
A hierarchy of testbed federation authorities govern the slices and projects, and assign security attributes to them.  Each slice is associated with one or more owners (PIDs).  The exemplary stitch policy maintains a set of authorized PIDs (a logical ACL) and approves a stitch request from any slice controlled by an authorized PID.   The ExoGENI provider API requires the ``hard-to-guess'' secret as a one-time passcode to validate mutual consent for cross-slice stitch requests~\cite{DCC17}.  A customer network (e.g., a campus) can also request a stitch at a named static stitchport with a named VLAN (network segment), if the NSP exposes static stitchports. 
% QQQ How does it know if a static stitchport request is authorized?  
% QQQ Can an ExoPlex NSP expose a static stitchport?
%There is no governance structure that binds a PID to the VLAN yet, and the NSP only takes the request from authorized users and assume that they do not steal other user's VLAN.

{\bf Connectivity}. Connectivity is off by default, and all flows are permissioned by policies of the endpoints.  We view edge NSPs as virtual software-defined exchanges (SDX) because they act as intermediaries to enforce declared policies of their customers.  Transit across an SDX is enabled only for flows that comply with applicable customer policies.  On the first packet of a previously unapproved source-destination pair, the SDX invokes a guard to authorize connectivity before installing SDN rules to pass traffic for the flow.
Exemplary connectivity policies are discussed in~\cite{CNERT18}.
The customer network is responsible to route outbound packets of a permissioned flow into the L2 link to the SDX. The SDX ensures that any inbound packets it routes onto the link are from permissioned flows.  The campus network simply delivers these packets to the destination subnet.

{\bf Routing}.  Traffic is also subject to origin authentication and route validation for secure routing, and endpoint path control policies to qualify all NSPs in the path.  NSPs apply related guard checks on both sides of API calls to propagate routes and routing policies.  The remaining subsections focus on these aspects.

\subsection{Origin Authentication}

Origin authentication ensures that the first advertisement of a prefix issues from a principal that is duly authorized to control routing to the prefix.  The origin must be valid according to statements issued by authority principals according to some trust structure.  The authority structure and logical checks ensure that endpoints communicate using non-conflicting IP prefixes and are prevented from stealing or controlling one another's traffic.

Our approach is analogous to the RFC 6480 architecture (RPKI), but implemented using SAFE.  The profiles for resource certificates are given by a logical vocabulary within the standard SAFE certificate format, validated by the logical rules in Listing~\ref{lst:own-prefix}.  We do not use special end-entity or Route Origination
Authorization (ROA) certificates as RPKI does; instead, any owner of a prefix may originate a route for the prefix to a provider network (e.g., an edge NSP or SDX).  The SAFE certificate store acts as the distributed repository system, but linked using SAFE's general hashed tokens.  In contrast, RPKI organizes stored certificates in a hierarchy, which is restrictive but also allows filesystem-like naming. 

The governance policy for prefix ownership identifies a set of one or more roots of authority for the address space, via local policy statements at each participating NSP that those principals are considered authoritative for specified prefixes and have the right to allocate sub-ranges from them.  One option models current IP governance as reflected in RPKI deployments: the local policy of each participant states that a root namespace authority (e.g., IANA/ICANN and its Internet Registries) controls all IP address space and allocates sub-ranges (prefixes) to owning principals hierarchically and transitively.  Participants must agree on the root authority and the form of the certified delegations, or else they fail to validate one another's prefixes.  An alternative is to ground prefix ownership in a forest of {\it a priori} anchors for disjoint segments of the IP address space.
This alternative is more practical for inter-domain networking on testbeds in that it does not rely on global authority deployment.
%non- conflicting IPv4 prefixes advertised to it by its customers

Listing~\ref{lst:allocate-prefix} shows the issuer script with the template for a resource certificate.  The certificate contains a logic statement to {\tt allocate} an IP prefix, i.e., to declare that a subject principal {\tt \$Holder} holds the prefix.  It also links to another certificate for support:  the token {\tt \$Cert} is the head of a chain of resource certificates grounded in some authority, and proving that the issuer owns an IP prefix containing the more specific {\tt \$Prefix} that it sub-allocates in this new resource certificate.   Anyone with this certificate may invoke a guard that fetches the chained certificates and validates the prefix ownership and that the chain is grounded in some locally accepted authority.  The guard validates by applying the logic rules in Listing~\ref{lst:own-prefix}.  We added a builtin operator ({\tt <:}) to the logic engine to validate containment of IPv4 prefixes specified in a standard string format.

\begin{lstlisting}[ caption = {A template script to post a statement of IP prefix allocation to anther principal, with a link to a certificate chain proving that the issuer controls the allocated IP prefix.}, label={lst:allocate-prefix}]
defcon ipAllocate(?Holder,?Prefix,?Cert) :-
  {
    link($Cert).
    allocate(?Holder,$Prefix).
  }.
\end{lstlisting}

\begin{lstlisting}[caption = {Logical rules to validate prefix ownership. A principal owns (holds) an IP prefix if an upstream issuer asserts that it does, and the issuer controls a containing prefix. ``$< :$'' is a boolean operator for containment of prefix values, built into the logic engine.  Its arguments are unified to prefix values before testing containment.}, label={lst:own-prefix}]
ownPrefix(?Holder,?Prefix):-
  $TrustRoot: allocate(?Holder,?Prefix).

ownPrefix(?Holder,?Prefix):-
  ?UpStream: allocate(?Holder,?Prefix),
  ownPrefix(?UpStream,?SupPrefix),
  ?Prefix <: ?SupPrefix.
\end{lstlisting}

\subsection{Route Validation}
\label{sec:route-validation}

%The inter-domain control plane for ExoPlex NSPs comprises new controller functions to propagate routes and policy notifications to peers.   They run in response to event notifications that trigger changes in routing.  In the prototype, NSP controllers generate event notifications by invoking REST service APIs of their peer NSP controllers.  

The NSP API in Table~\ref{tab:apis} allows peers and customers to attach to an NSP dataplane at specified peering points, register attached subnets, and enable traffic flows between prefix pairs.   After attaching, a neighbor may advertise a new prefix or routing policy at any time by passing a certificate link (token) through the API.   The receiving NSP validates it, integrates it with its routing base, installs SDN rules to implement it in its data plane, and propagates it to peer NSPs through their APIs. 

Each route advertisement is represented by a logic certificate. Listing~\ref{lst:originate-route} shows the template script for a customer network to originate a route for its IP prefix. It links its customized path control rules and the certificate for its allocated IP prefix.
Listing~\ref{lst:advertise-route} shows the template script for an NSP to sign an advertised route.
To propagate a route, an NSP invokes this script to issue an {\tt advertise} statement for each eligible peer, adding itself to the head of a sequence of NSPs on the path. It then invokes its peer's control API, passing the token to inform it of the route.  

\begin{lstlisting}[caption = {A template script for a customer network to originate a route advertisement with links to its customized routing policies and a certificate chain proving ownership of the advertised prefix.}, label={lst:originate-route}]
defcon originateRoute(?DstIP,?Path,?Target,
  ?IPCert):-
    ?Policy := label("custom policy"),
    ?NspAcl := label("nsp-tag-acl"),
  {
    link($IPCert).
    link($Policy).
    link(NspAcl).
    advertise($DstIP,$Path,$Target).
  }.
\end{lstlisting}

\begin{lstlisting}[caption = {A template script for an NSP to sign a route advertisement with a link to the signed statements of the previous hop and a link to a certificate chain that proves the tags of its network.}, label={lst:advertise-route}]
defcon advertiseRoute(?DstIP,?Path,?Target,
  ?Cert):-
    ?TagSubjectSet := label("tags"),
  {
    link($Cert).
    link($TagSubjectSet).
    advertise($DstIP,$Path,$Target).
  }.
\end{lstlisting}

Each route advertisement links to the certificate for the previous hop.
The links create chains of certificates, enabling standard logical rules to validate an entire path. 
List~\ref{lst:route-logic} shows the logic policy that an NSP enforces to verify a route advertisement.

\begin{lstlisting}[ caption = {\textbf{SAFE routing logic. The NSP($Self$) verifies a received advertisement by authorizing the route advertisement chain from the prefix owner. ($Path$ is a list of NSPs. $eq([?Head|?Tail], ?Path)$ is a built-in function that assigns the first element of $Path$ to $Head$ and the rest to $Tail$.}}, label={lst:route-logic}]
authorizedRoute(?Owner, ?DstIP, ?Path, ?AS):-
  eq([?Owner|?Tail], ?Path),
  eq(?Tail, []),
  ?Owner: advertise(?DstIP, ?Path, ?AS),
  ownPrefix(?Owner, ?DstIP).

authorizedRoute(?Owner, ?DstIP, ?Path, ?AS):-
  eq([?Head|?Tail], ?Path),
  ?Head:advertise(?DstIP, ?Path, ?AS),
  authorizedRoute(?Owner, ?DstIP, ?Tail, ?Head).
\end{lstlisting}

\begin{lstlisting}[ caption = {A  path control policy  for a  customer endpoint network.  
The  path  is compliant to the policy iff each hop of the path is authorized by the customer policy.}, 
label={lst:customer-policy}]
compliantPath(?SrcIP, ?DstIP, ?Path) :- 
  eq([?Head|?Tail], ?Path), 
  eq(?Tail, []),
  authorizedAS(?SrcIP, ?DstIP, ?Head).
  
compliantPath(?SrcIP, ?DstIP, ?Path) :- 
  eq([?Head|?Tail], ?Path),
  authorizedAS(?SrcIP, ?DstIP, ?Head), 
  compliantPath(?SrcIP, ?DstIP, ?Tail).
  
authorizedAS(?SrcIP, ?DstIP, ?AS):-
  nspTagAclEntry(?SrcIP, ?DstIP, ?Tag),
  tagAccess(?Tag, ?AS).
\end{lstlisting}

A subnet owner can specify path control rules for its traffic and link them to the route advertisement. Listing~\ref{lst:customer-policy} shows an example inbound path control policy that allows NSPs with specific tags to carry traffic from the source prefix to the subnet. 
Listing~\ref{lst:inbound-policy} shows a template script for a prefix owner to endorse NSPs with the specific tag to carry traffic from the source prefix to its prefix.

\begin{lstlisting}[ caption = {A template script for a customer network to specify required attributes of NSPs for its inbound traffic.}, label={lst:inbound-policy}]
defcon inboundPolicy(?Tag, ?Src, ?Dst) :-
  {                                 
    nspTagAclEntry($Tag, $Src, $Dst).
    label("nsp-tag-acl").
  }.
\end{lstlisting}

\subsection{Discussion: Governance}
\label{sec:governance}

As described in \S\ref{sec:secure-routing-overview}, the off-the-shelf rules for secure routing in ExoPlex depend on additional logic for governance policy.  A governance policy is a set of logical statements comprising facts to identify roots of authority and rules to validate certified delegations of authority from those roots.   Each participant executes code to configure its own policy by installing a set of logic statements.  Of course, interaction depends on compatible policies: the security logic blocks unsafe interactions that conflict with a participant's policy.

%Path control policies (\S\ref{sec:path-control}) also require a governance structure to label NSPs and subnets with secure attributes (tags).
 %For example, a customer policy may specify trust anchors for IP prefix ownership (e.g., IANA) and endorsement of qualifying NSPs; these trust roots might be imported via membership in a testbed federation such as GENI~\cite{geni07}.  %\note{redundant} 

Rather than specify a ``one size fits all'' governance structure, we use logical trust as a foundation to build and evolve governance structures and enable customers to specify their policy within those structures. 
Governance in the demonstration experiments works as follows.  Each principal has a keypair.  Slices and projects are objects approved by a controlling authority, which may make statements about them.

\begin{itemize}

\item A common federation root endorses a set of campus network authorities (CNA), one for each participating campus or enterprise.  The root also publishes a common governance policy (see below).

\item A common authority for the IP address space delegates disjoint prefixes to CNAs.   The CNAs delegate sub-prefixes to owners of edge subnets. 

\item Subnet owners control traffic on their subnets, e.g., they may install the prefix on a testbed slice and/or ``opt in'' to an edge NSP (an SDX) $S$ by issuing network bypass commands to stitch to $S$ at L2, register $S$ as the local gateway for traffic to selected prefixes, and accept traffic sourced from selected prefixes from $S$.

\item Once attached, subnet owners may advertise routes and publish path control policies to $S$.

\item Testbed slices have security metadata mirroring the idealized federation governance structure for GENI~\cite{brinn2015genitrust}.  NSP authorization policies for stitching may reference this metadata (\S\ref{sec:nsp-authz}). 

\item A common network authority endorses NSP public keys, binds them to slices, and asserts security attributes of NSPs for use in path control (\S\ref{sec:path-control}).  Local policies may accept other endorsing authorities on a per-attribute basis.
\end{itemize}

Any participant may validate compliance with its locally accepted governance policy.   Delegations result in chains or DAGs of linked logical certificates.   The governance rules represent safety predicates for these structures.  Any principal can apply its own rules to check validity for itself end-to-end, or delegate checking to another party, such as an edge NSP that acts as its provider.  Local governance policies are freely mobile: given a link to a policy of another principal, it is easy to import the policy's facts and rules and apply them.

This property of the logic system also makes it easy for a participant to delegate their governance policy to an authority.   In our demonstration prototype, all participants subscribe to a common package of governance logic posted by a common root of authority that is trusted by all parties.
The governance policy set installed by the local operator at each participant comprises simply a link to the remote policy set and a statement that accepts its conclusions: ``If the policy authority's rules conclude $P(x)$ then believe $P(x)$.'' 

%{\bf [Does an NSP controller ever propagate routes to a customer? NSP doesn't propagate route to customer network in our implementation.  But the route is initiated by customer networks. So we can still say this.]}

%In other work we are developing integration with existing identity management infrastructure and the policy support that is building out around that (e.g., inCommon Shibboleth with CoManage and CILogon).   We can extract trust information from a variety of sources and reason about it in a unified way using a common logic engine.   In ExoPlex, the logic engine checks compliance with logical access policies, e.g., for qualifying NSPs, permissioned peering, and permissioned flows.   

%% file: sections/experiment.tex
\label{sec:experiments}

The ExoPlex prototype is suitable for experiments with secure routing and policy flexibility involving modest numbers of customer prefixes and NSPs.  We conducted demonstration experiments on the ExoGENI testbed and I2-AL2S research fabric.   The performance and scale of these experiments are bounded primarily by the current limitations of the fabric, the VM-based OpenVSwitch routers we use on ExoGENI, and the Ryu SDN controllers for each NSP.  All compliance checks and crypto operations are off of the dataplane, and so affect only the setup times, and not the transit performance.  There is an obvious tradeoff between policy granularity and scale: fine-grained policies lead to fragmentation of the prefix space and routing/SDN flow tables, which could be a scaling barrier. 

We conducted several demonstration experiments based on the topology shown in Figure~\ref{fig:example-2},
%and a subset omitting NSPs 3 and 4.
%Figure~\ref{fig:example-1}.  
%These experiments interconnect eight and ten different ExoGENI slices respectively.
involving ten ExoGENI slices.
Each slice is instantiated under its own keypair and runs with its own controller, SDN, and SAFE logic engine, as in Figure~\ref{fig:exoplex}. 
%Additionally, each NSP/SDX slice has its own ExoPlex NSP controller and an accompanying SDN controller instance to manage its dataplane.
The customer networks A, B, C and D advertise subnet prefixes delegated to them through common governance authorities, which also endorse the networks with various security properties (tags), as described and shown in the figure.  The NSP controllers interact via REST APIs to peer and propagate edge-to-edge routes and policies.

%^/tree/chameleon/tree/tridentcom2019}.

\subsection{Experiment 1: Inbound Path Control}

\begin{figure}[htb!]
\begin{subfigure} [b]{0.45\textwidth}
\small{
table=0, n\_packets=1, ip, nw\_dst=192.168.10.0/24

table=0, n\_packets=1, ip, nw\_dst=192.168.30.0/24 
}
\caption{Flow table of $N1$.}
\end{subfigure}
%\par\bigskip
\begin{subfigure} [b]{0.45\textwidth}
\small{
 table=0, n\_packets=1, ip, nw\_dst=192.168.20.0/24 
 
 table=0, n\_packets=1, ip, nw\_dst=192.168.40.0/24
 }
\caption{Flow table of $N2$.}
\end{subfigure}

\if 0
\subfloat{%
  \includegraphics[clip,width=\textwidth]{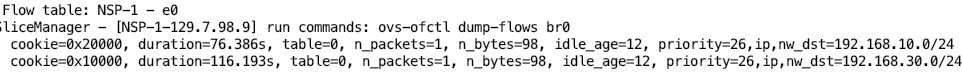}%
}
\subfloat{%
  \includegraphics[clip,width=\textwidth]{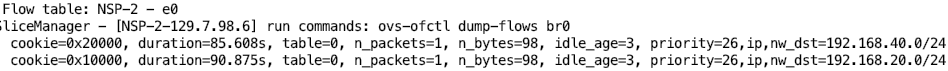}%
}
\fi
\caption{Flow tables of NSP switches in Experiment 1. We dump the NSP flow tables manually, filter the output flow entries by IP prefixes and packet counters for clarity, and omit unnecessary fields and the tables for $S1$ and $S2$. The packet counters confirm that traffic between subnets A and C traverses the path $\langle S1, N1, S2\rangle$ and traffic between subnets B and D takes the path $\langle S1, N2, S2 \rangle$. }
\label{fig:flow-ex1}
\end{figure}

\if 0
\begin{figure*}
\begin{center}
\includegraphics[width= 0.85\textwidth]{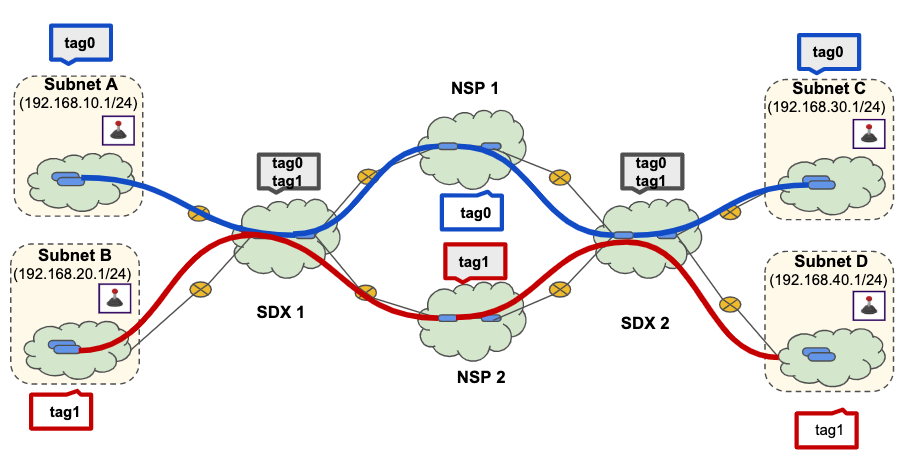}
\caption{ An exemplary inter-domain network used for Experiment 1. The customers have inbound path control policies that authorize only NSPs with compatible attributes to carry traffic to their subnets.}
\label{fig:example-1}
\end{center}
\end{figure*}

\fi

We evaluated inbound path control for the scenario shown in Figure~\ref{fig:example-2}, but omitting NSPs $N3$ and $N4$.
On ExoGENI it takes about 6.5 minutes to provision this topology and stitch the peering links, limited by provisioning times for I2-AL2S circuits.   The customer subnets stitch to their SDX providers concurrently,
%with multiple stitching requests to the same SDX  synchronized,
and then advertise their routes and path control policies when the stitches complete.  NSPs validate those advertisements and policies,  configure their dataplanes via SDN accordingly, and propagate them to their peers.  
%All of these actions occur asynchronously across the service networks.  NOTE: those operations (authorize, configure, propagate) are acutally sequential and should be sequential in my opinion
%The service networks (SDXs and NSPs) originate routes, authorize and propagate routes asynchronously. 

%The experiment has two pairs of customer subnets.  
For this experiment, customers A and C both authorize only the NSPs with secure attribute ``tag0'' to carry their inbound traffic, and authorize connectivity only with  edge subnets bearing the same attribute ``tag0''.   Customers B and D similarly authorize only ``tag1" for their network traffic.
%
%When subnet A stitches to $SDX 1$, it originates a route for A's prefix and includes a link for its inbound policy.
Upon receiving the route advertisement with linked inbound policy from A, $S1$ 
%checks whether A's inbound policy authorizes each of its neighbors $NSP 1$ and $NSP 2$ to carry traffic to A, and decides to 
propagates the route to $N1$ only, based on A's policy. 
%Upon receiving the advertisement, 
$N1$ validates the route, adds it to its cache of known routes, and propagates it to its authorized neighbor.
%and through to $SDX 2$.  
Other routes and policies are advertised and propagated throughout the network similarly.
%and chooses and installs the shortest known route,
%Each NSP's controller installs calls its network manager to install its chosen (shortest authorized) routes in its SDN network, while the SDX doesn't permit the flow unless the flow is allowed by the client connectivity policy.  
Then, customer pairs A-C and B-D each request a flow to the partner.
\if 0
Customer A then requests $S1$ to establish connectivity with customer C, i.e., to enable packets to flow between A and C in either direction.  
SDX $S1$ checks the request against the connectivity policies of customer C (linked with customer C's route advertisement). 
%{\color{red} Ingress filtering for incoming traffic at the edge SDX or egress filtering for outgoing traffic at the edge SDX, which makes more sense, or we can do both?}
If this requested connectivity is authorized, SDX $S1$ enables traffic for the permissioned flow.   Customer C makes a matching request to  $S2$ with a similar effect.  

%To filter out unauthorized traffic targeting client networks, both source and destination IP address blocks ought to be specified in the OpenFlow rules in each edge SDX.{\bf Still not quite clear about what is inferred from pre-placed policy on the connect request, and what is passed with the connect request, and what the blocks are.}
\fi

It takes about 3 seconds for each route advertisement to propagate throughout the network and enable flows between the pairs.
After these requests complete, we ping between subnet pairs with 1 packet and dump the flow tables from NSP switches, shown in Figure~\ref{fig:flow-ex1}, to verify that traffic follows the compliant paths as shown in Figure~\ref{fig:example-2}.
%As a result, customer A and customer C establish a communication channel via $<SDX 1, NSP 1, SDX2>$, and customer B and customer D communicate via $<SDX 1, NSP 2, SDX 2>$.

\subsection{Experiment 2: Outbound Path Control}
\ifdefined \CNERT
\input{sections/exp2-cnert}
\fi
\ifdefined \ARXIV
\input{sections/exp2-arxiv}

\fi

\subsection{SAFE Routing Authorization Performance}
We conducted experiments to evaluate the cost of logical authorizations in isolation.  We evaluated the inference performance of a SAFE server for validating routes and checking policy compliance under a throughput-limited synthetic workload.  The figure of merit is authorization ops per second (authz-ops/sec) for the checks performed by an NSP when it receives an advertisement. We run the SAFE server on a machine with 16 2.6 GHz cores (Intel Xeon E5-2650 v2) and saturate it with concurrent authorization queries through its REST API. The evaluation measures the cost to process the network calls and the cost to run the logic query on logic content extracted from a linked certificate DAG. 
%It includes checking the route for all steps along the route advertisement chain, checking prefix ownership by following prefix range delegations back to the governance root, and (optionally) checking compliance of each NSP in the path with inbound and/or outbound customer policies.

% and logic rules to validate a chain of such delegations.  The scripts link each range
%rom another machine at full rate to drive the SAFE server machine to full CPU utilization. 

We generated a synthetic topology of 1.8K NSP networks with a pattern of random peering among them: NSPs originate routes for their own IP prefixes and propagate routes to authorized neighbors randomly.  We also generated synthetic governance principals to delegate NSP security tags and customer IP prefixes from a common root.  As in the previous experiments, the customer path control policies query NSP attributes (tags) endorsed from the authorities.  We chose a tag delegation depth of 3 and IP prefix delegation depth of 3.

\begin{figure}[htb!]
\begin{center}
  \epsfig{file=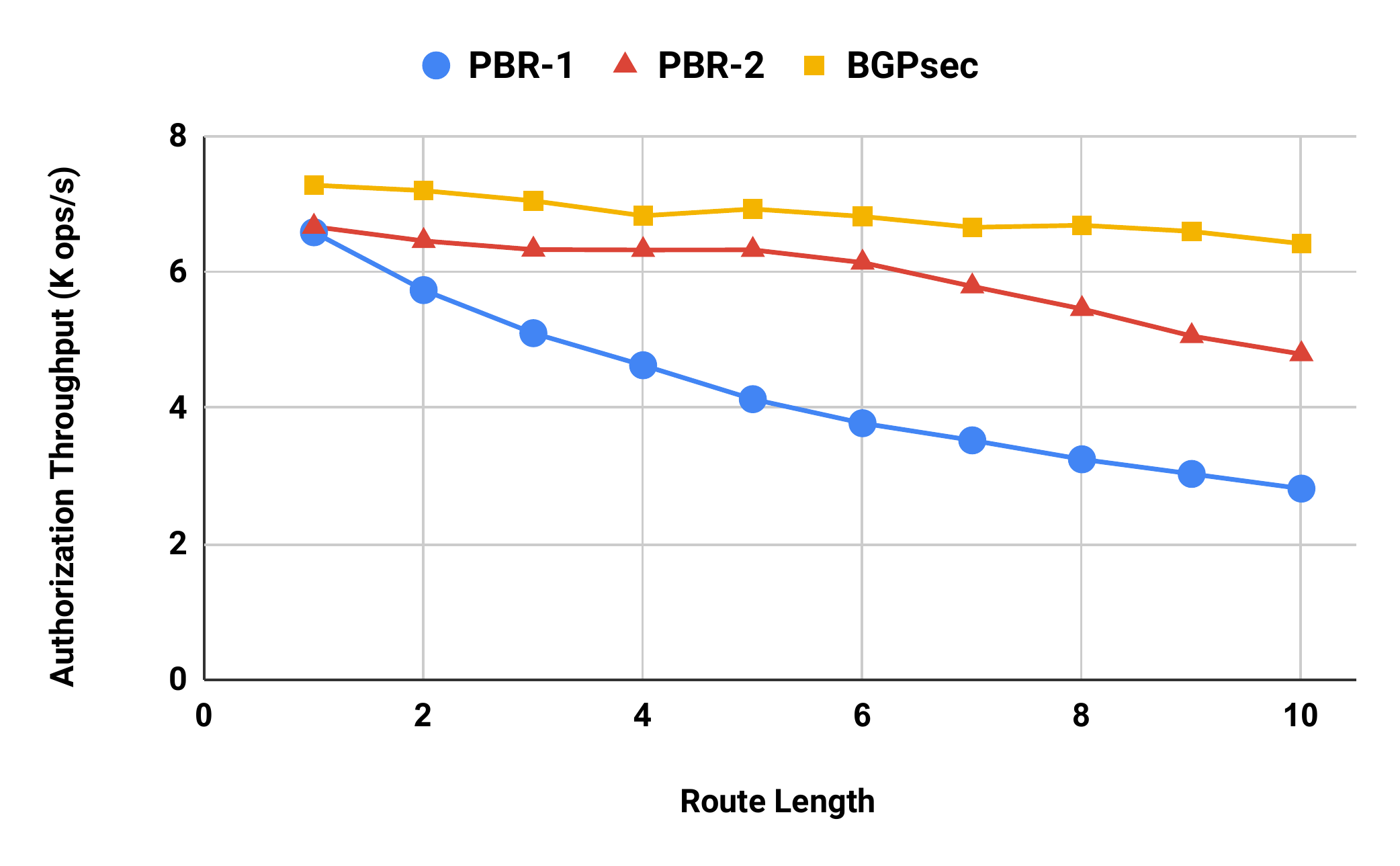, width=\figfactor\textwidth}
  \caption{SAFE logic engine throughput to validate routes with logical origin authentication and route attestation (equivalent to {\bf BGPsec}), with inbound path control policies ({\bf PBR-2}), and with both inbound and outbound path control ({\bf PBR-1}). 
  %These results include the logical inference cost alone, without certificate fetch latency or signature validation.
  }
  \label{fig:safe-bench}
\end{center}
\end{figure}

%The inference depth (the length of inference chain) for route authorization in SAFE routing is $9 + 10\cdot (L-1)$, where $L$ is the route length. 
Figure~\ref{fig:safe-bench} shows route authorization throughput as a function of route length for three sets of policies.  
There is a fixed cost to verify IP prefix ownership and validate routes, and an additional cost to check compliance with inbound and outbound path control policies at each NSP in the path.
%to check if the AS is compliant to the path control policies and the advertisement are authorized.
Thus the most expensive policy is {\bf PBR-1} (Policy-Based Routing), which checks both inbound and outbound policies.  {\bf PBR-2} checks inbound path control policy only.  
We compare the results to basic BGPsec-like route validation and prefix ownership alone without customer-specified path control (labeled as {\bf BGPsec}).
 % For each route advertisement, PBR-1 checks compliance with the source's outbound policy and the destination's inbound policy.  

 %{\bf BGPsec} includes origin authentication and route validation only: it authorizes the IP prefix ownership and the route advertisement chain in a manner equivalent to BGPsec, but specified using logical rules (e.g,, Figure~\ref{fig:safe-routing}) and logical certificates, and checked with the general-purpose SAFE engine.  

The results show that even for the most costly workload an NSP controller can check more than 2K routes per second.  These checks occur only when the NSP receives a new route or policy, and do not impact the dataplane.   While Figure~\ref{fig:safe-bench} does not include any crypto overhead (signature validation), these costs are fundamental for any routing security approach based on public-key cryptography (e.g., BGPsec).  SAFE does impose additional costs to fetch linked certificates on demand, but the SAFE engine validates them once and caches their logic content until the TTL expires, which minimizes these costs for policies, governance endorsements, etc.  (Figure~\ref{fig:safe-bench} ran on a hot cache pre-warmed with all relevant certificates.)  These results suggest that logical trust is fast enough to be practical at substantial scale.

%The throughput for route authorization in SAFE path control policy-based routing is less than that of BGPsec-style policies. Compared with BGPsec-style policies, SAFE routing introduced cost for authorizing AS based on customer path control policies for each AS along the route. This additional cost is dependent on customer path control policies and depths of privilege delegations. 

\ifdefined\ARXIV

\subsection{Prefix Pair Matching with AQT}
\label{sec:aqt-cost}
We implemented AQT in Java that supports prefix pair updates and queries. We randomly generated IP prefixes with prefix length 8, 16 and 24, and IP prefixes with longer prefix length are children of IP prefixes with shorter prefix length. We randomly generate different numbers of prefix pairs from those IP prefixes and default prefix ``0.0.0.0/0", with average prefix length about 23.8 and different average overlapping sizes. We  insert, query and delete all prefix pairs with AQT and measure the performance. The result is shown in Figure~\ref{fig:policymatching}. The results are consistent to the theoretical time complexity. 

\begin{figure}[t!]
\begin{center}
  \epsfig{file=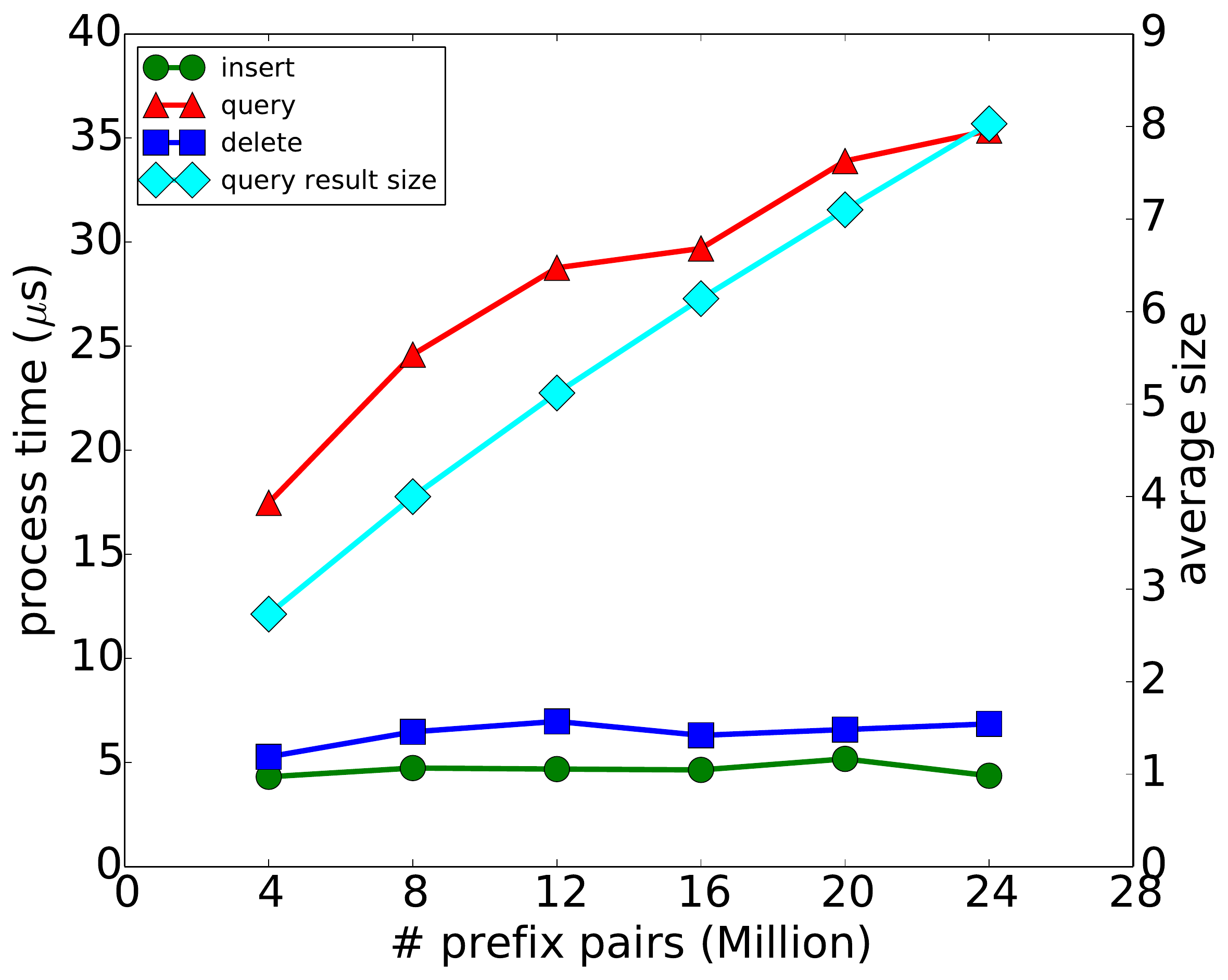, width=\figfactor\textwidth}
  \caption{Processing time for each prefix pair insertion, deletion and query operation with AQT. The time for insertion and deletion don't change much with the number of prefix pairs, as prefix lengths are about the same for all groups of prefix pairs. The time for query increases as the size of overlapped prefixes increases.}
  \label{fig:policymatching}
\end{center}
\end{figure}

\subsection{Dataplane Overhead for Source-Specific Routing}
\label{sec:source-specific}

We evaluate the overhead of source-specific routing with OpenVSwitch (OVS)~\cite{pfaff2015ovs}. 
OVS runs the virtual switches in the ExoGENI NSP deployments for our experiments.  The exemplary policies for connectivity and path control in this paper require SDN routing rules that match packets on both source and destination.  For example, with outbound path control policies, packets to a given destination $D$ might take different routes depending on the source $S$.  This fine-grained policy control of routing may inflate routing tables and complicate packet classification at the SDN layer.  The purpose of this section is to explain and quantify this cost, using the OVS software switch as a reference point.

OVS manages OpenFlow tables in userspace and a microflow cache and a megaflow cache in kernel. The megaflow cache is a single table of disjoint entries that caches the recent flows.  The megaflow cache has a capacity of 200,000 flow entries in the latest mainstream OVS versions. The number of rules in megaflow cache are related with both the number of OpenFlow entries and the number of active flows. Packets in cached microflows or megaflows are fast processed in kernel. When the number of OpenFlow entries and active flows are both large, the megaflow cache will overflow and there will be cache misses. OVS will match the missed packet with OpenFlow tables in userspace, which is slow and expensive. 

OVS classfies packets in both OpenFlow tables and the megaflow table with tuple space search. A tuple is a set of fields matched in the flow entries. Flow entries that match on the same tuple are put in the same hash table, where the keys are the hashes of the matched fields. The cost of tuple space search depends on the number of unique tuples (i.e. the number of hash tables) specified in the flow entries. 

We evaluate the CPU overhead of packet classification with OpenFlow entries that match on both source and destination IP addresses in OVS. We create an ExoGENI slice with three ``XO Extra Large'' nodes in linear topology on the same rack working as source node, OVS node and sink node respectively.
We run Open vSwitch 2.12.0 in Ubuntu 19.10 on the second node with 4 2.2 GHz cores (Intel(R) Xeon(R) CPU E5-2660 v2). We evaluate nine sets of policies. For the three sets of destination-based OpenFlow entries({\it dst entries}), we randomly generate OpenFlow entries with the destination prefix within 32.0.0.0/8 and the length of the netmask between 8 and 24. The distribution of generated prefix lengths are subject to the IPv4 prefix cumulative distribution in~\cite{bgpin2018}. There is also a default destination based OpenFlow entry that matches on ``32.0.0.0/8" with lowest priority. For source-based OpenFlow entries({\it src entries}), we randomly generate entries that match on both source and destination IPv4 prefix with source IP prefixes in 16.0.0.0/8 and the same destination IP prefixes as in {\it dst entries}.
For each set of {\it dst entries}, we add $1x$ and $64x$ {\it src entries} additionally, resulting in six sets of {\it mixed entries}. We generate 8 {\it pcap} files with different number of tcp packets and with empty payload(66 Byte/packet) whose source and destination IP addresses are uniform randomly distributed in 16.0.0.0/8 and 32.0.0.0/8.

With this worst-case synthetic traffic mix, the number of flows identified by source and destination IP address is about the same as the number of packets. We replay the traffic with {\it{tcpreplay}} at fixed rate 50K packets per second for multiple rounds and collect total CPU utilization of the OVS VM in a 10-minute period.  We also collect how many packets are received on sink node. The CPU utilization in the first round is higher than following rounds, as OVS needs to process every new flow with userspace OpenFlow tables. We omitted CPU utilization results in the first few rounds to illustrate the impact of source-specific routing for long-lived flows. We also sample and counter megaflow entries with {\it ovs-appctl} in separate runs.

The CPU cost for packet classification is related to the number of OpenFlow entries, the type of the entries and the number of flows.
Compared with {\it dst entries}, {\it src entries} leads to more packet classification cost due to two major factors. First,
there are more hash tables for each packet to be matched against. Second, the number of flow entries could be much larger with {\it src entries}. Given limited capacity, the megaflow cache for {\it src entries} overflows more easily, increasing the costs to fall back to tables in userspace.
 
 Figure~\ref{fig:cpu-mega} shows the CPU utilization and the number of sampled megaflows with different numbers of flows and different numbers and types of OpenFlow entries. 
 The CPU utilization with {\it mixed entries} is higher than that with {\it dst entries}.
 With a small number of flows or a small number of OpenFlow entries, the packets are processed in the fast path with the cached microflows and megaflows in kernel for both {\it dst entries} and {\it mixed entries}. 
 And the CPU utilization for {\it mixed entries} are about $1x$ to $20x$ higher than that for {\it dst entries} with less than 200,000 flows. 
 When the megaflow cache is not full, the number of OpenFlow entries has limited impact on CPU utilization, as the classification cost mainly depends on the number of unique tuples in megaflow table. 
 However, the number of OpenFlow entries still matters: the number of megaflows with more OpenFlow entries is also larger. 
 The megaflow cache for more/larger OpenFlow entries overflows earlier as the number of flows increases, as we can see from the Figure~\ref{fig:cpu-mega}.
With more than 200,000 flows and a large number of {\it mixed entries}, the megaflow cache overflows, generating much higher processing overhead in userspace. The CPU utilization with {\it mixed entries} could be more than $50x$ higher. In our experiment, the maximum number of megaflows for {\it dst entries} is capped at $6.5K$ as the destination IP address are sampled from a relatively small prefix 32.0.0.0/8. Therefore, we the megaflow cache for {\it dst entries} does not overflow, and the CPU utilization for {\it dst entries} is low. But with {\it dst entries} specified in a larger address space, we also expect to see the performance drop with a large number of OpenFlow entries and flows.

%Source-specific routing may contribute processing overhead in OVS. First, it results in more tuples in both OpenFlow tables and megaflow tables, making the packet classification more expensive. Second, there will be more flows in megaflow cache, resulting in more cache misses and processing overhead in userspace when the cache becomes full. Assuming that the prefix lengths are between 8 and 24, source-specific routing inflates the number of hash tables by at most 16 times.

%\begin{figure}[htb]
%\begin{subfigure} [b]{0.5\textwidth}
%\centering
%  \epsfig{file=figs/flow-cpu-1.pdf, width=\textwidth}
%  \caption{The CPU utilization with different number of flows and policies.}
  %These results include the logical inference cost alone, without certificate fetch latency or signature validation.
% \label{fig:flow-cpu}
% \end{subfigure}
% \par\bigskip
% \begin{subfigure} [b]{0.5\textwidth}
%  \epsfig{file=figs/flow-mega.pdf, width=\textwidth}
 % \caption{The number of megaflows sampled during the replay.}
  %These results include the logical inference cost alone, without certificate fetch latency or signature validation.
% \label{fig:flow-mega}
 %\end{subfigure}
%\end{figure}

\begin{figure}[htb]
\centering
  \epsfig{file=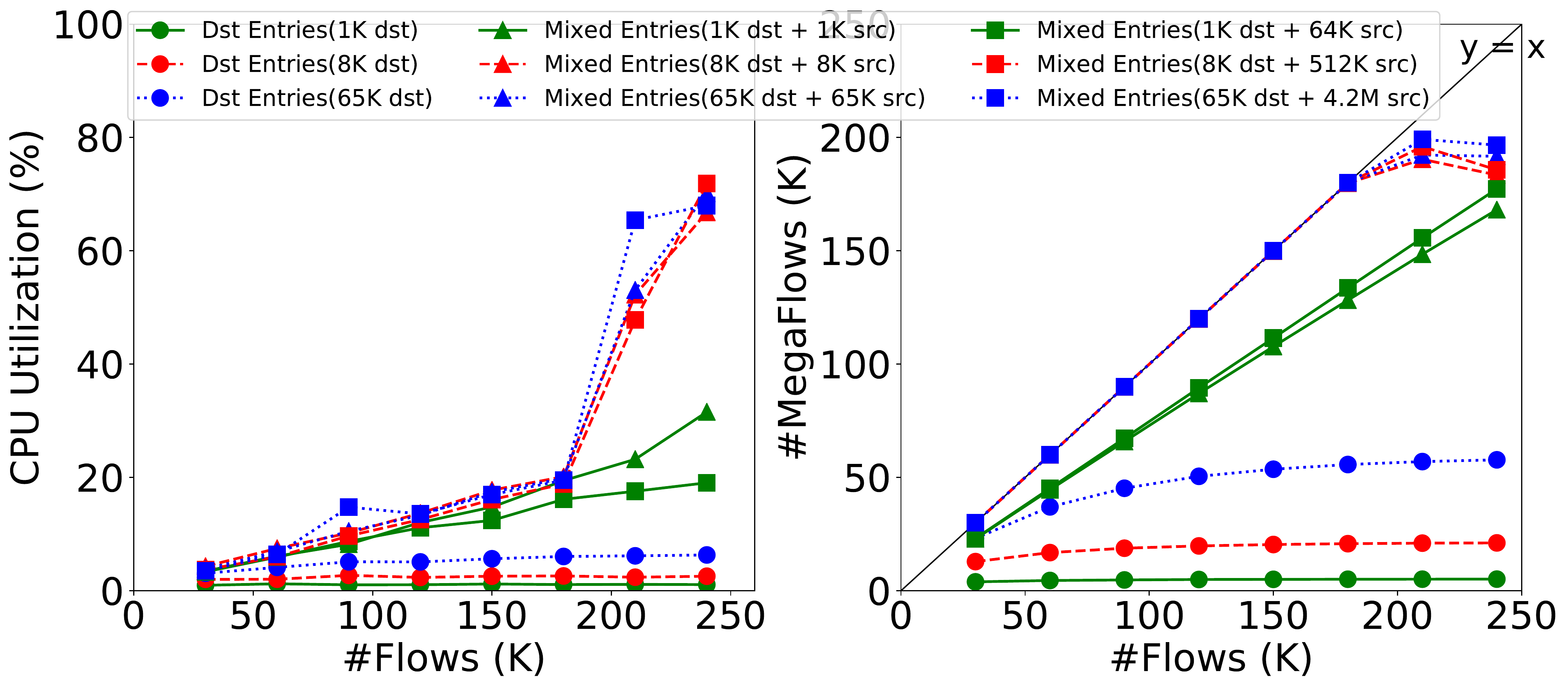, width=0.49\textwidth}
  \caption{The CPU utilization and the number of sampled megaflows with different numbers of flows and different numbers and types of OpenFlow entries. The packet loss rates are all less than 2\%.}
  %These results include the logical inference cost alone, without certificate fetch latency or signature validation.
 \label{fig:cpu-mega}
\end{figure}
\fi

%% file: sections/exp2-cnert.tex
For this experiment we extended the inter-domain network with NSPs $N3$ and $N4$, as shown in Figure~\ref{fig:example-2}, and added outbound policies.   NSPs propagate route advertisements through the network to all NSPs that are compatible with both the inbound and outbound policies and have a route to the destination.  It takes about 10 minutes to provision the topology and 30 seconds for routes to propagate and stabilize.

%There are four client networks, two SDX networks and four NSP networks. The authorities has delegated IP prefixes and secure tags to all those networks, as shown in the figure. The connectivity policies of these clients are that only connections with other client networks with the same secure tags are allowed.

%Before attaching to a provider, each client posts an {\it inbound path control} policy that restricts the NSPs that may carry its incoming traffic.  In addition, clients may optionally post policy rules for {\it outbound path control} that limit the NSPs for its outgoing flows.
%Both types of rules may optionally specify different policies to apply to flows with specific IP prefixes.   

%{\bf  [Or clients with specified attributes?]{\color{red} This is also possible, but this will results in a lot of work when matching source policies and route advertisements. We have to run SAFE queries for each pair to see if it is a match. By specifying source and destination IP address, we can just match on the IP prefixes, which is fast}} 
%clients use APIs at the provider SDX to advertise these policies.   

The customer subnets again constrain their inbound traffic to transit through NSPs with compatible tags.  Customers A and C have additional inbound and outbound  policies that only NSPs with secure tag ``tag0'' can carry traffic between subnet A (192.168.10.0/24) and subnet C (192.168.30.0/24).  Similarly, customers B and D restrict the traffic between their subnets to pass through NSPs with secure tag ``tag1'', and A and D restrict their flows to NSPs with ``tag2''.

After the connections are set up, we ping between pairs of subnets and inspect the flow tables (not shown) as before to verify that traffic takes the expected compliant paths.

%% file: sections/exp2-arxiv.tex
For this experiment we extended the inter-domain network with NSPs $N3$ and $N4$, as shown in Figure~\ref{fig:example-2},
and added outbound policies. The routing policies in our experiment settings are shown in Table~\ref{tab:route-policy}.

\begin{table}[htb]
\begin{center}
\begin{tabular}{
|p{1cm}|p{2.5cm}|p{2.5cm}|  }
 \hline
 \multirow{2}{4em}{Subnet Owner}&\multicolumn{2}{|c|}{Inbound/Outbound Policies} \\\cline{2-3}

  &Src/Dst Subnet&Required Tags\\
 \hline
\multirow{2}{4em}{A} & C & tag0\\\cline{2-3}
& D & tag2\\
 \hline
B & D & tag0, tag1\\
\hline
B & C (only at step 6)& tag0\\
\hline
C & A & tag0, tag1 \\
\hline
\multirow{2}{4em}{D} & A & tag2\\\cline{2-3}
& B & tag1\\
\hline
\end{tabular}
\end{center}
\caption{The inbound and outbound path control policies of subnets in Figure~\ref{fig:example-2}. In this setting, the inbound and outbound path control policies of a subnet are symmetric (but not necessarily symmetric), i.e., they require the same set of security tags for NSPs. The route between two subnets must be compliant to the path control policies of both the source subnet and destination subnet. For example, ``tag 0'' is the only legal security tag for routes between subnet A and C.}
\label{tab:route-policy}
\end{table}

\begin{figure}[htb!]
\begin{center}
\includegraphics[width=0.48 \textwidth]{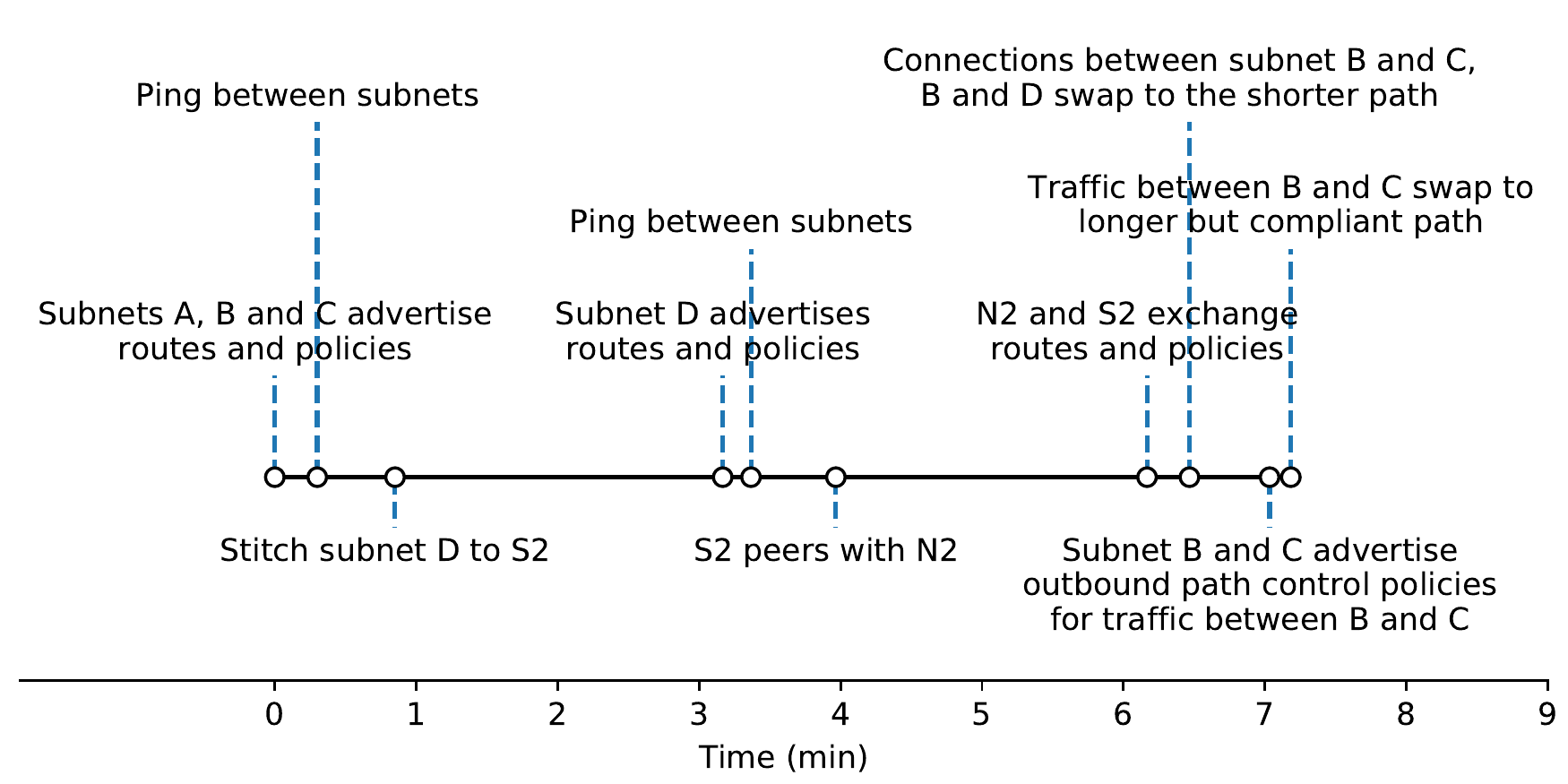}
\caption{Timeline of experiment 2.}
\label{fig:timeline}
\end{center}
\end{figure}

We carry out the experiment in the following steps: (1) Stitch all NSPs. (2) Stitch subnets A, B and C to its NSPs. (3) Subnets A, B and  C advertise routes with both source and destination address specified and their outbound policies. The NSPs authorize and propagates those routes and policies and make dataplane configurations accordingly. (4) Stitch subnet D to $S2$ and advertise its routes and policies. (5) Stitch $S2$ and $N2$ directly to provide a shorter routes for subnet B to reach subnet C and D. (6) Subnet B and C advertise inbound and outbound path control policies that require ``tag0" for traffic between their subnets. It takes about 12 minutes to stitch all NSPs and subnets A, B and C. Figure~\ref{fig:timeline} shows a timeline of the experiment starting at step (3). 

\begin{figure}[htb]
\begin{subfigure} [b]{0.45\textwidth}
\small{

n\_packets=4,nw\_src=192.168.10.0/24,nw\_dst=192.168.30.0/24 

n\_packets=4,nw\_src=192.168.30.0/24,nw\_dst=192.168.10.0/24

n\_packets=3,nw\_src=192.168.40.0/24,nw\_dst=192.168.10.0/24 

n\_packets=3,nw\_src=192.168.10.0/24,nw\_dst=192.168.40.0/24 

n\_packets=1,nw\_src=192.168.20.0/24,nw\_dst=192.168.30.0/24

n\_packets=1,nw\_src=192.168.30.0/24,nw\_dst=192.168.20.0/24
}
\caption{Flow table of $N1$.}
\end{subfigure}
\par\bigskip
\begin{subfigure} [b]{0.45\textwidth}
\small{
n\_packets=3,nw\_src=192.168.40.0/24,nw\_dst=192.168.20.0/24

n\_packets=3,nw\_src=192.168.20.0/24,nw\_dst=192.168.40.0/24

n\_packets=3,nw\_dst=192.168.20.0/24

n\_packets=3,nw\_dst=192.168.30.0/24
}
\caption{Flow table of $N2$.}
\end{subfigure}
\par\bigskip
\begin{subfigure} [b]{0.45\textwidth}
\small{
n\_packets=4,nw\_src=192.168.10.0/24,nw\_dst=192.168.30.0/24

n\_packets=4,nw\_src=192.168.30.0/24,nw\_dst=192.168.10.0/24 

n\_packets=1,nw\_src=192.168.30.0/24,nw\_dst=192.168.20.0/24
 
n\_packets=1,nw\_src=192.168.20.0/24,nw\_dst=192.168.30.0/24
}
\caption{Flow table of $N3$.}
\end{subfigure}
\par\bigskip
\begin{subfigure} [b]{0.45\textwidth}
\small{
n\_packets=3,nw\_src=192.168.40.0/24,nw\_dst=192.168.10.0/24

n\_packets=1,nw\_src=192.168.40.0/24,nw\_dst=192.168.20.0/24

n\_packets=1,nw\_src=192.168.20.0/24,nw\_dst=192.168.40.0/24 

n\_packets=3,nw\_src=192.168.10.0/24,nw\_dst=192.168.40.0/24

n\_packets=2,nw\_dst=192.168.20.0/24

n\_packets=2,nw\_dst=192.168.30.0/24
}
\caption{Flow table of $N4$.}
\end{subfigure}
\caption{Flow tables of NSP switches in Experiment 2 after step 6.}
\label{fig:flowtable-2}
\end{figure}

% Traffic between subnet A and subnet C and between A and D traverse the paths $\langle S1, N1, N3, S2\rangle$ and $\langle S1, N1, N4, S2\rangle$ respectively. And traffic between B and D take the paths $\langle S1, N2, N4, S2\rangle$ before step 5 and $\langle S1, N2, S2\rangle$ after step 5. Traffic between subnet B and subnet C is subject to only their default inbound path control policies, thus any path is compliant in this example. In our experiment, it takes the paths $\langle S1, N2, N4, S2\rangle$ before step 5, $\langle S1, N2, S2\rangle$ at step 5 and $\langle S1, N1, N3, S2\rangle$ at step 6.

We try sending a packet between connection pairs A and C, A and D, B and C, as well as B and D each time after step 3, 4, 5 and 6. 
Table~\ref{tab:flow-path} shows the paths that different flows take at different steps. The paths are compliant to the path control policies of the subnets as expected. 
Figure~\ref{fig:flowtable-2} shows the flow tables of NSP switches after step 6.
The packet counters of the flow tables proves the correctness of Table~\ref{tab:flow-path}.
Before step 6, traffic between subnet B and C are subject to the default policies that require ``tag0" or ``tag1" for the NSPs. In our run, traffic between subnet B and C took the route $\langle S1, N2, N4, S2\rangle$ before step 5 and $\langle S1, N2, S2\rangle$ when the shorter route became available at step 5. After step 6, only the route $\langle S1, N1, N3, S2 \rangle$ is compliant for traffic between subnet B and C. We verify the actual route that those packets take by checking the packet counters of the flow tables of the NSPs after each step. 

\begin{table}[htb]
\begin{center}
\begin{tabular}{
|p{0.6cm}|p{0.4cm}|p{2cm}| p{1.6cm}|p{2cm}| }
 \hline
 Flow&3&4&5 &6\\
\hline
A\ding{214}C&\multicolumn{4}{|c|}{$\langle S1, N1, N3, S2\rangle$}\\
\hline
A\ding{214}D&&\multicolumn{3}{|c|}{$\langle S1, N1, N4, S2\rangle$}\\
\hline
B\ding{214}C&\multicolumn{2}{|c|}{$\langle S1, N2, N4, S2\rangle$}&$\langle S1, N2, S2\rangle$&$\langle S1, N1, N3, S2\rangle$\\

\hline
B\ding{214}D&&$\langle S1,N2, N4,S2\rangle$& \multicolumn{2}{|c|}{$\langle S1, N2, S2\rangle$}\\
\hline
\end{tabular}
\end{center}
\caption{The path for traffic between subnet pairs at each step.}
\label{tab:flow-path}
\end{table}

%% file: sections/conclusion.tex
We propose a logical trust approach to network security for testbed-hosted Network Service Providers (NSPs), implemented in the ExoPlex network controller platform, and extend it for secure policy-based routing for {\it interdomain} networks with multiple NSPs.  ExoPlex can be the basis for a ``testbed for trust'' for inter-domain networks that are constructed on the fly and span multiple slices, testbeds, and campuses.  NSP owners and customers may experiment with policy for peering, routing, path control, and governance by specifying custom policies in logic, without changing the ExoPlex code.  In particular, the trust plane supports customer policies for permissioning the NSPs themselves, so that customer traffic does not pass through untrusted NSPs (path control).  A secure foundation with at least these features is a necessary prerequisite for safe testbed opt-in by real customer traffic---an important aspirational goal.  

{\bf Acknowledgment.}  This material is based upon work supported by the US National Science Foundation (NSF) under Grants No. (ACI-1642140, ACI-1642142, CNS-1330659, CNS-1243315) and through the Global Environment for Network Innovations (GENI) program.  Any opinions, findings, and conclusions or recommendations do not necessarily reflect the views of NSF.

%The initial experiments show how to use ExoPlex as a platform for multi-domain network experiments.

%and manage multi-domain authorizations with trust logic (SAFE). 

%The strategies for adapting the service network to meet dynamic network requirements remain to be explored.